\documentclass[proc]{edpsmath}

\usepackage{stmaryrd}
\usepackage{times}
 \usepackage[LY1]{fontenc}
\usepackage{enumerate,cancel} 
\usepackage{xspace}
\usepackage{savesym,ulem,amsmath}
\def\em{\it} 
\savesymbol{subequations}
\usepackage{cases}

\usepackage{ifthen,bm}
\usepackage{dsfont}
\usepackage{pgf}
\usepackage{epsfig,graphicx,epstopdf}
\usepackage{latexsym, amsfonts, amssymb}
\usepackage{alltt,latexsym, verbatim}
\usepackage{float}
\usepackage{amsfonts}
\usepackage{xspace}
\usepackage{natbib}\def\citeN{\citet*}\def\shortciteN{\citet}
\newtheorem{theorem}{Theorem}[section]  

\newtheorem{conj}{Conjecture}[section]

\numberwithin{equation}{section}

\newcommand{\bt}{\begin{theorem}}\newcommand{\et}{\end{theorem}}
\newcommand{\bl}{\begin{lmm}}\newcommand{\el}{\end{lmm}}
\newcommand{\bp}{\begin{proposition}}\newcommand{\ep}{\end{proposition}}
\newcommand{\bcor}{\begin{corollary}}\newcommand{\ecor}{\end{corollary}}
\newcommand{\bconj}{\begin{conj}}\newcommand{\econj}{\end{conj}}

\newcommand{\bd}{\begin{definition} \rm }\newcommand{\ed}{\end{definition} }
\newcommand{\brem }{\begin{rmrk} \rm }\newcommand{\erem }{\end{rmrk}}
\newcommand{\bex}{\begin{example} \rm }\newcommand{\eex}{\end{example}}

\def\proof{\noindent \textbf{\emph{\textbf{Proof}.$\qqq$}}}
\def\finproof {\hfill $\Box$ \vskip 5 pt }

\def \be{\begin{eqnarray}}
\def \ee{\end{eqnarray}}
\def \b*{\begin{eqnarray*}}
\def \e*{\end{eqnarray*}}

\def \CC{{\mathbb C}}
\def \DD{{\mathbb D}}

\def \LL{{\mathbb L}}
\def \MM{{\mathbb M}}
\def \NN{{\mathbb N}}
\def \PP{{\mathbb P}}

\def \Z{{\mathbf{Z}}}

\def \[{[\,\!\![}
\def \]{]\,\!\!]}

\def \1{{\bf 1}}

\def \proof{{\noindent \bf Proof. }}

\def\F{{\cal T}}\def\F{{\cal G}}\def\F{{\cal F}}

\newcommand{\bea}{\begin{eqnarray*}}
\newcommand{\eea}{\end{eqnarray*}}
\newcommand{\beqa}{\begin{eqnarray}}
\newcommand{\eeqa}{\end{eqnarray}}

\def\N{{\mathbb N}}\def\N{N}
\def\R{{\mathbb R}}
\def\proof{\noindent {\it Proof. $\, $}}
\def\finproof {\hfill $\Box$ \vskip 5 pt }
\def\I{\mathds{1}}
\def\sp{\,,\ \,}\def\sp{,\ \,}

\def\r{\eqref}

\def\bal{\begin{aligned}}
\def\eal{\end{aligned}}

\def\hat{\widehat}

\def\p{p}
\def\mym{m}

\def\them{m}

\def\xitau2{\xi_{(\thetau)}}\def\xitau2{\xi}
\def\xiitau2{\xi^i_{(\thetau)}}\def\xiitau2{\xi^i}
\def\xintau2{\tilde{\xi}_{(\thetau)}}\def\xintau2{\tilde{\xi}}

\def\Ltau2{\tilde{\xi}(\tau_{\mym},\tau_1)}
\def\chiitau2{P^i_{\tau}}

\newcommand{\beq}{\begin{eqnarray*}}
\newcommand{\eeq}{\end{eqnarray*}}

\def\mynu{\nu}

\def\mym{-}

\def\theg{g}

\def\emph{}

\def\ds{\ud s}

\def\gg{{\mathbb G}}\def\gg{{\mathbb F}}
\def\G{{\cal G}}\def\G{{\cal F}}
\def\ff{\mathbb{F}}

\def\ff{\mathbb{F}^*}
\def\F{{\cal F}^*}

\def\gg{{\cal G}}\def\gg{\mathbb{G}}
\def\G{{\cal G}}
\def\ff{{\cal F}}\def\ff{\mathbb{F}}
\def\F{{\cal F}}
\def\hh{{\cal H}}

\def\E{\mathbb{E}}

\def\ta{\tm}

\def\Rf{\theOm}\def\Rf{\mathfrak{r}}
\def\Rf{\theOm}\def\Rf{\mathfrak{r}}\def\Rf{\mathfrak{r}^b}

\def\thisR{R}\def\thisR{\rho}\def\thisR{R}\def\thisR{\rho^b}

\def\thetau{\vartheta}

\def\tb{{\bar{\tau}}}\def\tb{\thetau}

\def\theomega{\nu}\def\theomega{\tau}\def\theomega{\tau}

\def\tb{\thetau}\def\tb{{\bar{\tau}}}
\def\theomega{\nu}\def\theomega{\tau}

\def\ind{\mathds{1}}
\def\rc#1{\textcolor[rgb]{0.98,0.0,0.0}{#1}}\def\rc{}
\def\e{z}\def\e{e}

\def\theN{\mathbb{N}_n}\def\theN{N}

\def\theC{\mathbf{C}}\def\theC{\boldsymbol{\vm }}\def\theC{{\Ms }}

\newcommand{\beql}[1]{\beqa\label{#1}\begin{aligned}}
\newcommand{\eeql}{\eal\eeqa}
\newcommand{\bel}{\bea\bal}
\newcommand{\eel}{\end{aligned}\eea}

\def\eee{\end{document}}
\def\ind {1\!\!1}\def\ind{\mathds{1}}
\def\I{\ind}

\def\F{{\cal F}}

\def\G{{\cal G}}

\def\ff{{\mathbb F}}
\def\hh{{\mathbb H}}

\def\gg{{\mathbb G}}

\def\P{\mathbb P}
 \def\Q{\mathbb Q}

\def\E {{\mathbb E} }

\def\mr{\textcolor{red}}
\def\mb{\textcolor{blue}}

\def\R{{\mathbb R}}
\def\N{{\mathbb N}}

\def\finproof {\hfill $\Box$ \vskip 5 pt }

\def\bal{\begin{aligned}}
\def\eal{\end{aligned}}

\def\finproof {\hfill $\Box$ \vskip 5 pt }

\def\Proba{\Proba}\def\Proba{\mathbb{Q}}

\def\sp{,\ \, }
\def\r#1{(\ref{#1})}

\def\xitau2{\xi_{(\thetau)}}
\def\xiitau2{\xi^i_{(\thetau)}}
\def\xintau2{\tilde{\xi}_{(\thetau)}}
\def\Ltau2{\tilde{\xi}(\tau_0,\tau_1)}
\def\chiitau2{P^i_{\tau}}

\def\thetau{\tau}

\def\ta{\tau_{\!-\!1}}\def\ta{\tau_{B}}
\def\thekappa{\kappa}\def\thekappa{S}

\def\cB{\mathcal{B}}

\def\cF{\mathcal{F}}
\def\cG{\mathcal{G}}
\def\cH{\mathcal{H}}

\def\cJ{\mathcal{J}}

\def\cZ{\mathcal{Z}}

\def\emph{}

\def\tb{{\bar{\tau}}}
\def\thisR{\thisR }

\def\rc{R_c}

\def\eee{\end{document}}
\def\textsl{}

\def\proof{\noindent {\it {\textbf{Proof}}}.$\;\,$}
\def\finproof {\hfill $\Box$ \vskip 5 pt }\def\finproof {$\Box$}\def\finproof{\rule{4pt}{6pt}}

\def \be{\begin{eqnarray}}
\def \ee{\end{eqnarray}}
\def \b*{\begin{eqnarray*}}
\def \e*{\end{eqnarray*}}

\def \CC{{\mathbb C}}
\def \DD{{\mathbb D}}

\def \LL{{\mathbb L}}
\def \MM{{\mathbb M}}
\def \NN{{\mathbb N}}
\def \PP{{\mathbb P}}

\def \Z{{\mathbf{Z}}}

\def \[{[\,\!\![}
\def \]{]\,\!\!]}

\def \1{{\bf 1}}

\def\F{{\cal T}}\def\F{{\cal G}}\def\F{{\cal F}}

\def\N{{\mathbb N}}\def\N{N}
\def\R{{\mathbb R}}

\def\I{\mathds{1}}

\def\bal{\begin{aligned}}
\def\eal{\end{aligned}}

\def\hat{\widehat}

\def\p{p}
\def\mym{m}

\def\them{m}

\def\xitau2{\xi_{(\thetau)}}\def\xitau2{\xi}
\def\xiitau2{\xi^i_{(\thetau)}}\def\xiitau2{\xi^i}
\def\xintau2{\tilde{\xi}_{(\thetau)}}\def\xintau2{\tilde{\xi}}

\def\Ltau2{\tilde{\xi}(\tau_{\mym},\tau_1)}
\def\chiitau2{P^i_{\tau}}

\def\mynu{\nu}\def\mynu{g}\def\mynu{f}\def\mynu{\alpha}

\def\mym{-}

\def\theg{g}

\def\tilde{\widetilde}
\def\overline{\bar}

\def\emph{}

\def\G{{\cal G}}\def\G{{\cal F}}
\def\ff{\mathbb{F}}

\def\ff{\mathbb{F}^{\star}}
\def\F{{\cal F}^{\star}}

\def\gg{{\cal G}}\def\gg{\mathbb{G}}
\def\G{{\cal G}}
\def\ff{{\cal F}}\def\ff{\mathbb{F}}
\def\F{{\cal F}}
\def\hh{{\cal H}}\def\hh{\mathbb{H}}

\def\E{\mathbb{E}}

\def\ta{\tm}

\def\Rf{\theOm}\def\Rf{\Rf}\def\Rf{R_{f}}\def\Rf{\bar{R}_{b}}

\def\thisR{R}\def\thisR{\thisR }\def\thisR{R_b}

\def\thetau{\vartheta}

\def\tb{{\bar{\tau}}}\def\tb{\thetau}

\def\tb{\thetau}\def\tb{{\bar{\tau}}}

\def\paragraph{\noindent\textbf}

\def\theomega{\tau}

\def\qqq{\quad\quad\quad}

\def\theN{N}

\def\eee{\bibliographystyle{chicago}\bibliography{B}\end{document}} \def\eee{\end{document}}
\def\themu{\mu}\def\themu{\eta}\def\themu{\beta}

\def\bthet{\boldsymbol\theta}\def\bthet{\mathbf{k}}

\def\db{\tb^\thisdelta }

\def\tt{\tilde{\theta}}\def\tt{\theta^{\star}}\def\tt{\xi^\delta_{\star}}

\def\gt{\cb_{\hat{\tau}}}

\def\qr#1{\eqref{#1}}

\newcommand{\indi}[1]{\I_{\{{#1}\}}}
\newcommand{\iend}{\end{itemize}}
\newcommand{\ok}{\rule{4pt}{6pt}}\renewcommand{\ok}{\finproof}
\newcommand{\desb}{\begin{description}}
\newcommand{\dese}{\end{description}}

\newcommand{\dcb}{\begin{array}{lll}}
\newcommand{\dce}{\end{array}}
\newcommand{\ebe}{\begin{enumerate}[1)]\setlength{\baselineskip}{13pt}\setlength{\parskip}{5pt}}
\newcommand{\dbe}{\end{enumerate}}

\newcommand{\ibegin}{\begin{itemize}\setlength{\baselineskip}{19pt}\setlength{\parskip}{7pt}}

\def\cro#1{\langle #1\rangle}

\makeatletter
\newenvironment{systeme*}{\arraycolsep=1.4pt\left\{\begin{array}{l}}{\end{array}\right.}

\makeatother

\def\om{\omega}

\def\cro#1{{\langle#1\rangle}}

\def\bc{c}

\def\subset{\subseteq}

\def\thexit{\xi^{\star}}\def\thexit{\xi_{\star}}\def\thexit{\bar{\xi}_{\tau}}

\def\theLambda{(1-\Rf)}\def\theLambda{\Lambda}

\def\B{B}

\def\r{\textcolor{red}}
\def\b{\textcolor{blue}}

\def\eq{:=}\def\eq{=}
\begin{document}

%%-----------------------------
%%      the top matter
%%-----------------------------
\title{Invariance Properties in the Dynamic Gaussian Copula Model}\thanks{This research benefited from the support of the 
``Chair Markets in Transition'',  
F\'ed\'eration Bancaire Fran\c caise, and of the ANR project 11-LABX-0019. }% At most 5 thanks
\author{St\'{e}phane Cr\'{e}pey}\address{Universit\'e d'\'Evry Val d'Essonne, Laboratoire de Math\'ematiques et Mod\'elisation d'\'Evry and UMR CNRS 8071, 91037 \'Evry Cedex, France.}
\author{Shiqi Song}\address{Universit\'e d'\'Evry Val d'Essonne, Laboratoire de Math\'ematiques et Mod\'elisation d'\'Evry and UMR CNRS 8071, 91037 \'Evry Cedex, France.} 
%
%\dedicated{\it Dedicated to Maurice Dupont} %if necessary
%
\begin{abstract} We prove that the default times (or any of their minima)
%$\tau_i$ 
in the dynamic Gaussian copula model of \citeN{CrepeyJeanblancWu12} are invariance times
in the sense of \citeN{CrepeySong15c}, with 
related invariance probability measures different from the pricing measure.  
This reflects a departure from the immersion property, whereby the default intensities of the surviving names and therefore the value of credit protection  
spike at default times. These properties are in line with the wrong-way risk feature
of counterparty risk embedded in credit derivatives, i.e.~the adverse dependence between the default risk of a counterparty and an underlying credit derivative exposure. \end{abstract}
%
%\begin{resume} ... \end{resume}
 
\maketitle

\textbf{Keywords:}
counterparty credit risk,
wrong-way risk, Gaussian copula, 
dynamic copula, immersion property, invariance time, CDS.

\vspace{2mm}
\noindent
\textbf{Mathematics Subject Classification:} 
91G40, %Credit risk. 
60G07.% Probability theory and stochastic processes
%60G44, %Martingales with continuous parameter

%\tableofcontents

\section{Introduction}\label{s:intr}

This paper deals with the mathematics of the 
dynamic Gaussian copula (DGC) model of
\citeN{CrepeyJeanblancWu12} (see also \citeN[Chapter 7]{BieleckiBrigoCrepeyHerbertsson13} and \citeN{CrepeyNguyen15}).
As developed in \shortciteN[Section 7.3.3]{BieleckiBrigoCrepeyHerbertsson13}, this model yields a dynamic meaning to
%Greeks in this model are consistent with 
the ad hoc bump sensitivities that were used by traders for hedging CDO tranches by CDS contracts before the subprime crisis. From a more topical perspective, it can be used for counterparty risk computations on CDS portfolios.
%, a feature commonly found in financial softwares. 
Related models include
the one-period Merton model of \citeN[Section 6]{FermanianVigneron11} or other variants commonly used in credit and counterparty risk softwares.

The dynamic Gaussian copula  model has been assessed from an engineering perspective in previous work, but a detailed mathematical study, including explicit computation of the main model primitives, has been deferred to the present paper.

\subsection{Invariance Times and Probability Measures}\label{ss:inva}
 
We work on a filtered probability space $(\Omega, \gg, \mathcal{A},\mathbb{Q})$. 
Given a $\gg$ stopping time $\tau$ and a subfiltration $\ff$ of $\gg,$ $\ff$ and $\gg$ satisfying the usual conditions,
let $J$ and $S$ denote the survival indicator process of $\tau$ and its optional projection known as the Az\'ema supermartingale of $\tau,$ i.e.~\index{s@$S$}
$$J_t=\indi{\tau>t}\sp S_t={^{o}\!J}_t=\Q(\tau>t\,|\,\F_t)\sp t\geq 0.$$ 

The following conditions are studied in 
\citet{Crepey13a1,CrepeySong15c}.\\
%\citeANP{Crepey13a1} (\citeyearNP{Crepey13a1},\citeyearNP{CrepeySong15c}).\\

\noindent
\index{b@(B)} 
\noindent\textbf{Condition (B)}. 
Any $\gg$ predictable process $U$ admits an $\ff$ predictable reduction, i.e.~an
$\ff$ predictable process, denoted by $U^\prime,$ that coincides with ${U}$ on 
$\rrbracket 0,\tau \rrbracket$.\\
%\vspace{5pt}

\noindent
For any left-limited process $Y,$ 
we denote by
$Y^{\theomega-}=JY +(1-J)Y_{\tau-}$ 
 the process $Y$ stopped before $\tau$.\\

\index{c@(A)} 
\noindent\textbf{Condition (A)}. Given a constant time horizon $T>0$,
there exists a probability measure $\mathbb{P}$
equivalent to $\mathbb{Q}$
on {${\cal F}_{T}$}
%, with $\P$ expectation denoted by \index{e@$\tilde{\E}$}$\tilde{\E},$
such that $({\ff},\mathbb{P})$ local martingales
{stopped before $\tau$}
are $(\gg,\mathbb{Q})$ local martingales on $[0,T]$.\vspace{5pt}

If the conditions (B) and (A) are satisfied, then we say that $\tau$ is an invariance time and $\mathbb{P}$ is an invariance probability measure.
%The conditions (B) and (A) provide more flexibility than a standard progressive enlargement setup, by allowing for a full model filtration $\gg$ larger than $\ff$ progressively enlarged by $\tau$ and by allowing playing with the measure $\P.$ 
If, in addition, $S_T>0$ almost surely,
then $\ff$ predictable reductions are uniquely defined on $( 0,T]$ and any inequality between two $\gg$ predictable processes on $( 0,\tau]$ implies the same inequality between their $\ff$ predictable reductions on $(0,T]$ (see \citeN[Lemma 6.1]{Song14}); invariance probability measures are uniquely defined on $\cF_T,$ so that one can talk of the invariance probability measure $\P$ (as the specification of an invariance probability measure outside $\cF_T$ is immaterial anyway).

{\def\e{i}
\def\mys{0}
\def\bb{\mathbb{B}}
\def\B{\mathcal{B}}

\def\kI{\mbox{supp}(\bthet_t)}\def\kI{\mathfrak{I}}\def\kI{\mathcal{I}}
\def\kIs{\mbox{supp}(\tilde{\bthet}_t)}\def\kIs{\mathfrak{I}^\star}\def\kIs{\mathcal{I}^\star}
\def\BIs{\mathcal{B}^\star} 
\def\kJ{{\mathfrak{J}}}\def\kJ{{\mathcal{J}}} 
\def\kJs{\mathfrak{J}^\star}\def\kJs{\mathcal{J}^\star}
\def\kJt{\tilde{\mathfrak{J}}} \def\kJt{\tilde{\mathcal{J}}} 

\def\thexit{\hat{\xi}}\def\thexit{\xi^{\star}}

\def\bal{\begin{aligned}}
\def\eal{\end{aligned}}

\def\la{\label}
\def\bbf{\mathbf}
\def\al{\alpha}
\def\beq{\begin{eqnarray}}
\def\eeq{\end{eqnarray}}
\def\La{\theLambda}
\def\si{\sigma}\def\si{\mynu}
\def\sommen{\sum_{j,k\in V_N}}
\def\sommezd{\sum_{k\in\bbZ^d}}
\def\eps{\varepsilon}
\def\what{$\clubsuit\clubsuit\clubsuit$}
\def\grad{\bigtriangledown}
\def\demonstration{\noindent{\textbf{Proof. }}}
\def\sss{\subset\subset}
\def\trace{\,{\rm Tr}\,}
\def\Om{\Omega}
\def\om{\omega}
\def\gb{\overline{g}}
\def\gc{\hat{g}}

\def\PP{\mathbb P}\def\PP{\Proba}

\def\Ne{\N^{\ast}}
\def\CC{{\cal C}}
\def\DD{{\cal D}}
\def\B{\mathcal{B}}
\def\SSS{{\cal S}}
\def\E{{\mathbb E}}
\def\hh{{\mathbb H}}
\def\F{{\cal B}}\def\F{{\cal F}}
\def\G{{\cal G}}
\def\cH{{\cal H}}
\def\HHl{{\cal H}_{loc}}
\def\II{{\cal I}}
\def\JJ{{\cal J}}
\def\KK{{\cal K}}
\def\LL{{\cal L}}
\def\ft{\tilde{f}}
\def\tf{\tilde{f}}
\def\tth{\tilde{h}}
\def\tg{\tilde{g}}
\def\gt{\tilde{g}}
\def\et{\tilde{E}}
\def\ut{\tilde{u}}
\def\phit{\tilde{\Phi}}
\def\hf{\hat{f}}
\def\hg{\hat{g}}
\def\NN{{\cal N}}
\def\MM{{\cal M}}
\def\OO{{\cal O}}
\def\BP{{ \overline{{\cal P}}}}
\def\R{\mathbb{R}}
\def\D{\mathbb{D}}
\def\dis{\displaystyle}
\def\Z{\mathbb{Z}}
\def\C{\mathbb{C}}
\def\ind{\mathds{1}}
\def\rgt{\rightarrow}
\def\p{^{\prime}}
\def\vphi{\varphi}
\def\si{\varsigma}\def\si{\sigma}\def\si{\mynu}
\def\pa{\partial}
\def\para{\parallel}
\def\ds{D}
\def\gr{\sigma^{\ast}\nabla u}
\def\db{d{{B}}_s}
\def\dbt{d{B}_t}
\def\HD{F}
\def\disp{\displaystyle}
\def\Es{(E^{\ast})}
\def\dpr{d^{\prime}}
\def\hu{\hat{u}}
\def\ve{\varepsilon}
\def\un{ {\bf 1}}

\def\them{m}\def\them{\mathbf{m}}
\def\thel{l}
\def\myGt{G}
\def\thel{l}
\def\theCt{C}
\def\theS{\kappa}\def\theS{S}
\def\theSl{\kappa_l}\def\theSl{S_l}
\def\thekappa{\kappa}
%% dongli's macro
\def\hC{\hat{C}}
\def\tC{\tilde{C}}
\def\Q{\mathbb{Q}}
\def\tZ{\tilde{Z}}
\def\hZ{\hat{Z}}
\def\ff{\mathbb{B}}\def\ff{\mathbb{F}}
\def\ta{\tilde{a}}
\def\ii{\imath}
\def\jj{\jmath}

\def\ir{\textcolor{red}}
\def\mr{\textcolor{green}}
\def\mb{\textcolor{blue}}
\def\rc{\textcolor{red}}
\def\mr{\textcolor{red}}
\def\mb{\textcolor{blue}}

\def\bal{\begin{aligned}}
\def\eal{\end{aligned}}
\def\thel{l}
\def\myGt{G}
\def\thel{l}
\def\theCt{C}
\def\theS{\kappa}\def\theS{S}
\def\theSl{\kappa_l}\def\theSl{S_l}

\def\hat{\widehat}

\def\thel{\ell}
\def\az{a} 
\def\thel{\ell}\def\thel{k}
\def\ind {1\!\!1}\def\ind{\mathds{1}}
\def\I{\ind}
\def \PP{{\mathbb Q}}
\def\theN{N}
\def\qqq{\quad\quad\quad}
\def\Phis{\Phi^\star}\def\Phis{\Phi}
\def\dl{\dot{\Psi}}\def\dl{\boldsymbol\psi^{i}}\def\dl{\psi^{i}}\def\dl{\psi^{j}}
\def\dli{\psi^{i}}
\def\z{\mathbf{z}}
\def\Z{\mathbf{Z}}
\def\cro#1{\langle #1\rangle}
\def\cZ{\mathcal{Z}}
\def\cZt{\tilde{\mathcal{Z}}}
\def\cZs{\mathcal{Z}^\star}
\def\Zm{\widetilde{m}}\def\Zm{\zeta}
\def\thel{l}
\def\thisc{c}
\def\theC{C}
\def\themu{\eta}\def\themu{\beta}

\section{Dynamic Gaussian Copula Model}
\label{s:dgc}

%\subsection{Outline of the Paper}\label{ss:outl}

In the paper we prove that, given a constant time horizon $T>0,$ the default times $\tau_i$ (or any of their minima)
in the DGC model are invariance times, with 
related invariance probability measures $\P$ uniquely defined and not equal to  
$\Q$ on $\cF_T.$ 
This reflects a departure from the immersion property, whereby the default intensities of the surviving names and therefore the value of credit protection  
spike at default times, as observed in practice.
%consistent with empirical facts. 
%This is in line with the wrong-way risk feature
%of counterparty risk embedded in credit derivatives, i.e.~the adverse dependence between the default risk of a counterparty and the underlying credit derivative exposure.
This feature makes the DGC model appropriate for dealing with
counterparty risk on credit derivatives (notably, portfolios of CDS contracts) traded between a bank and its counterparty, respectively labeled as $-1$ and $0,$ 
and referencing credit names 1 to $n,$ for some positive integer $n.$ 
Accordingly, 
we introduce $$
\index{n@$\N$}
\N=\{-1,0,1,\ldots,n\} \mbox{ and }
\index{n@$\N^\star$}
\N^\star =\{1,\ldots,n\}  
$$ 
and we focus on $\tau=\tau_{-1}\wedge \tau_0$ in the paper.
However, analog properties hold for any minimum of the $\tau_i$ and, in particular, for the $\tau_i$ themselves.

\subsection{The model}

We consider a family of independent standard linear Brownian motions $Z$ and $Z^i, i\in N$. For \index{r@$\varrho$}$\varrho\in [0,1)$, we define
\beql{e:BM}
B^i_t\eq \sqrt{\varrho} Z_t+\sqrt{1-\varrho} {Z}^i_t.
\eeql
Let \index{s@$\varsigma$}$\varsigma$ be a continuous function on $\mathbb{R}_{+}$ with $\int_{\mathbb{R}_{+}}\varsigma^2(s)ds=1$ and $\mynu^2(t)  \eq{\int_t^{+\infty} \varsigma ^2(s)ds}>0$ for all $t\in\R_+$. For any $i\in N,$ let \index{h@$h_i$}$h_i $ be a continuously differentiable
strictly increasing function from $\mathbb{R}_+^*$ to $\mathbb{R},$ with derivative denoted by $\dot{h}_i$,
such that $\lim_{
s\downarrow 0} h_i(s) = -\infty$ and $\lim_{s\uparrow+\infty}h_i (s)=+\infty$.
We define
\begin{equation}\label{e:tau}
 \tau_i\eq h_i^{-1}\big(\int_0^{+\infty}\varsigma(u)dB_u^i\big)
=
h_i^{-1}\big(\sqrt{\varrho}\int_0^{+\infty}\varsigma(u)dZ_u+\sqrt{1-\varrho}\int_0^{+\infty}\varsigma(u)dZ^i_u\big), 
\end{equation}
for $i\in N$. The random times $(\tau_i)_{i\in N}$ follow the standard one-factor Gaussian copula model of \citeN{Li00} (a DGC model in abbreviation),
with correlation parameter $\varrho$ and
with marginal survival function
$\Phi \circ h_i$ of $\tau_i,$
where 
$$
\Phi(t)\eq  \int_{t}^{+\infty}\frac{1}{\sqrt{2\pi}}e^{-\frac{x^2}{2}}dx, \ t\in\mathbb{R} 
$$
is the standard normal
survival function.
Note that, if $\varrho<1$, the $\tau_i$ avoid each other:$$
\mathbb{Q}(\tau_{i}=\tau_{j})=0, \mbox{ for any } i\neq j \mbox{ in } N .
$$ 

\subsection{Density Property}

{\def\gtime{\tau}
By multivariate density default model, we mean a model
with an $\ff$ conditional density of the default times (see e.g. the condition (DH) in \citeN[page 1800]{Pham}), given some reference subfiltration $\ff$ of $\gg$. This is
the multivariate extension of the notion of a density time, first introduced in an initial enlargement setup in
\citeN{Jacod87} and revisited in a progressive enlargement setup in
\citeN{JeanblancLeCam09b} (under the name of initial time) and
\citet*{ElKarouiJeanblancJiao10,ElKarouiJeanblancYiao15,ElKarouiJeanblancYiao09}.
% \citeANP{ElKarouiJeanblancJiao10} (\citeyearNP{ElKarouiJeanblancJiao10},\citeyearNP{ElKarouiJeanblancYiao15},\citeyearNP{ElKarouiJeanblancYiao09}).

First
we prove that the DGC model is a multivariate density model with respect to the natural filtration $\mathbb{B}=(\mathcal{B}_{t})_{t\geq 0}$ of the Brownian motions $Z$ and $Z^i, i\in N$. We introduce the following processes.
\bel
\index{m@$m^i$}
m_t^i\eq \int_0^t\varsigma(u)dB_u^i
\ \mbox{ and } \
\overline{m}_t^i\eq \int_t^\infty\varsigma(u)dB_u^i
=
h(\tau_{i}) - m^i_{t},\ i\in \N.
\eel 
%Recall that $\mynu^2(t)= {\int_t^{+\infty} \varsigma ^2(v)dv}$ is assumed positive for all $t\geq 0$. 
The standard normal density function is denoted by $$
\phi(x)\eq \frac{1}{\sqrt{2\pi}}e^{-\frac{x^2}{2}}, \ x\in \mathbb{R}.
$$

\begin{theorem}\label{l:explic} 
The dynamic Gaussian copula model is a multivariate density model of default times (with respect to the filtration $\mathbb{B}$),
with conditional Lebesgue density $$
p_t(t_i , i\in N)\eq  \partial_{t_{-1}}\ldots\partial_{t_n}\mathbb{Q}(\tau_i <t_i ,\, i\in N\,|\,\cB_t)
$$ 
of the $\tau_i,\, i\in N$, 
given,
for any nonnegative $t_i,$ $i\in N$, and $t\in\R_+,$ by
\beql{QoverSa} 
&p_t(t_i ,\, i\in N) 
=	
\int_\mathbb{R}\phi(y)\prod_{i\in N}\phi\left(\frac{h_i(t_i)-m^i_t+\si(t)\sqrt{\varrho}y}{\si(t)\sqrt{1-\varrho}}\right)\frac{\dot{h}_i(t_i)}{\si(t)\sqrt{1-\varrho}}dy.\eeql
\end{theorem}

\proof The conditional density function $p$ given $\cB_t$ can be computed thanks to the independence of increments of the processes $Z, Z^i, i\in N$. Actually, for any $t\geq 0$, we can write
$$
\tau_{i}
=
h^{-1}_{i}(m^i_{t} + \sqrt{\varrho}\xi + \sqrt{1-\varrho}\xi_{i}), \ i\in \theN,
$$
where $\xi$ is a real normal random variable with variance $\alpha^2_{t}$, where $(\xi_j)_{j\in N}$ is a
%an $|N|$ dimensional 
centered
Gaussian vector independent of $\xi$ with homogeneous marginal variances $\alpha^2_{t}$ and
zero pairwise correlations, 
and where the family $\xi, \xi_{i}, i\in N,$ is independent of $\mathcal{B}_{t}$. 
See \shortciteN[page 172]{BieleckiBrigoCrepeyHerbertsson13}\footnote{Or \shortciteN[page 3]{CrepeyJeanblancWu12} in the journal version.}. ~\finproof

\subsection{Computation of the intensity processes}
Note that the $\tau_i$ are $\mathcal{B}_{\infty}$ measurable, but they are not $\mathbb{B}$ stopping times. In the DGC model,
the full
model filtration $\gg=(\cG_t)_{t\ge 0}$ is taken as
the progressive enlargement of the Brownian filtration $\mathbb{B}$ by the
$\tau_i, i\in N$, augmented so as to satisfy the usual conditions, i.e.
\beql{e:pr}
\G_t= \cap_{s>t}(\mathcal{B}_{s} \vee \bigvee_{i\in \N} \sigma(\tau_i
 \wedge s)),\ \ t\geq 0.\eeql
In this section we prove that the $\tau_i$ are totally inaccessible $\gg$ stopping times with intensities that we compute explicitly.

For $I\subset \theN$ and $j\in \theN,$
we define: 
\bel 
&\rho^I
=\frac{\varrho}{|I|\varrho+1}, \quad (\sigma^I) ^2
=(1-\varrho)\frac{|I|\varrho+1}{|I|\varrho+1-\varrho},
\quad 
\lambda^I=\frac{\varrho}{(|I|-1)\varrho+1},
\\
&Z_t^{j,I}(u)
=\frac{h_j(u)-{m_t^j}}{\si (t)}- 
\lambda^I
\sum_{i\in I}\frac{\overline{m}_t^i}{\si (t)}.
\eel
For $t\geq 0$,
let$$
\kI_t = \{i\in N: \tau_{i}\leq t\}
$$ 
(representing the set of obligors in $N$ 
that are in default at time $t$) and let
\bel
&
{\rho}_t = {\rho}^{\kI_t},\
\sigma_t = \sigma^{\kI_t},\
\cJ_t= N\setminus\kI_t.
\eel
For $\sigma>0,$ $\rho\in[0,1]$ and $J\subset N$, we define the functions 
\beql{eqa initial enlargement} 
&\Phi_{J,{\rho},\sigma}(\z_{J})\eq \Q(\xi_j>z_j, j\in J)\sp
\dl_{J,\rho,\sigma}\big(\z_{J}\big)\eq -\frac{\partial_{z_j}\Phi_{J,\rho,\sigma}}{ \Phi_{J,\rho,\sigma} }\big(\z_{J}\big),\ j\in J, 
\eeql 
where 
%$d=|J|$,  
$\z_{J}=(z_j)_{j\in J} $ is a real vector and $(\xi_j)_{j\in N}$ is a %$(n+2)$ dimensional 
centered
Gaussian vector with homogeneous marginal variances $\sigma^2$ and pairwise
correlations $\rho.$ 
Note the following:
\bl\label{l:Gauss}
For $I=N\setminus J$, the family of random variables $$
\left(\xi_{j} - \frac{\rho}{(|I|-1)\rho + 1}\sum_{i\in I}\xi_{i}\right)_{j\in J}
$$
defines a centered Gaussian vector independent of $\sigma(\xi_{i}, i\in I)$, 
with
homogeneous marginal variances and pairwise
correlations, 
respectively given as
\begin{equation}
\sigma^2(1 - \rho)\frac{|I|\rho +1}{|I|\rho +1 - \rho}\ \ \mbox{ and }\ \
\frac{\rho}{|I|\rho +1}.~\finproof
\end{equation} 
\el

\bl For $u>0$, 
\begin{equation}\label{e:BPhi}
\mathbb{E}[\ind_{\{u_{j}<\tau_j, j\in J\}}|\mathcal{B}_t\vee\sigma(\boldsymbol{\tau}_I)]
=
\Phi_{J, \rho^I,\sigma^I}(Z^{j,I}_t(u_{j}), j\in J).
\end{equation}
\el
\proof For $j\in J$ and $u_{j}\in\mathbb{R}$, the condition $u_{j}<\tau_{j}$ is equivalent to
\begin{equation}
Z_t^{j,I}(u_{j})
=
\frac{h_{j}(u_{j}) - m^j_{t}}{\alpha(t)} - \lambda^I \sum_{i\in I}\frac{\overline{m}^i_{t}}{\alpha(t)}\ \
<\ \
 \frac{\overline{m}^j_{t}}{\alpha(t)} - \lambda^I \sum_{i\in I}\frac{\overline{m}^i_{t}}{\alpha(t)}
\end{equation}
Noting that $m^{j}_{t}\in \mathcal{B}_t, \overline{m}^i_{t}\in \mathcal{B}_t\vee\sigma(\boldsymbol{\tau}_I), i\in I$, the desired result follows by an application of Lemma \ref{l:Gauss}.~\finproof\\

\bl
For every $t>0$ and $I\subset N$ we have, writing
$J=N\setminus I$ and $\boldsymbol{\tau}_I=(\tau_{i})_{i\in I}:$
%(the sets $I,J$ being considered with their natural order), 
\beql{Gtsplit} 
&&\{\tau_i\leq t<\tau_j: i\in I, j\in J\}\cap\mathcal{G}_t=
\{\tau_i\leq t<\tau_j: i\in I, j\in J\}\cap(\mathcal{B}_t\vee\sigma(\boldsymbol{\tau}_I)).
\eeql
%where $J=N\setminus I$ and $\boldsymbol{\tau}_I$ denotes the vector $(\tau_{i})_{i\in I}$.~\finproof
\el

\proof
Let the $\gtime _{(i)}$ be the increasing ordering of the $\gtime _i,$ with also $\gtime _{(0)} = 0$ 
and $\gtime _{(n+1)}= \infty.$
According to the optional splitting formula which holds in any multivariate density
model of default times (see \citeN{Song12}),
for any $\mathbb{G}$ optional process $Y$, there exists a $\mathcal{O}(\mathbb{B})\otimes\mathcal{B}([0,\infty]^n)$-measurable functions $Y^{(i)}\sp i\in N,$ 
such that 
\beql{e:osf}
Y=\sum_{i=0}^n Y^{(i)}(\gtime_{-1}\nmid\gtime _{(i)},\ldots,\gtime _n\nmid\gtime _{(i)})\ind_{[\gtime _{(i)},\gtime _{(i+1)})},
\eeql
where $a\nmid b$ denotes $a$ if $a\leq b$ and $\infty$ if $a>b$, for $a,b\in[0,\infty].$  
Since $\cG_t=\sigma(Y_t)$ and 
$Y^{(i)}(\gtime _1\nmid\gtime _{(i)},\ldots,\gtime _n\nmid\gtime _{(i)})\ind_{[\gtime _{(i)},\gtime _{(i+1)})}$
is a function of 
$\mathcal{B}_t$ and $\boldsymbol{\tau}_I$
on
$\{\tau_i\leq t<\tau_j: i\in I, j\in J\},$
this implies \qr{Gtsplit}.
\finproof

\begin{theorem}
For any $j\in N,$ $\tau_j$ admits a $(\mathbb{G},\mathbb{Q})$ intensity given by 
\begin{eqnarray}
 \label{QoverSci} 
&&
\gamma^j_{t}
\eq 
\ind_{\{t<\tau_j \}}
 \frac{\dot{h} _j(t)}{\alpha(t)} 
 \dl_{\mathcal{J}_{t},\rho_t,\sigma_t}\big( Z^{j, \mathcal{I}_{t}}_t(t), j\in \mathcal{J}_{t} \big)\sp    t\in\R_+. 
\end{eqnarray}
\end{theorem}

\proof
 Let $l\in N$. For bounded $\mathcal{B}_t$ measurable functions $F$, for measurable bounded function $f$, for $0\leq t\leq s<\infty$, we look at$$
\mathbb{E}[Ff(\boldsymbol{\tau}_I)\ind_{\{\tau_i\leq t<\tau_j: i\in I, j\in J\}}\ind_{\{t<\tau_{l}\leq s\}}].
$$
We need only to consider $l \in J$. Then, using \qr{e:BPhi} to pass to the third line and conditioning in conjunction with the tower rule to pass to the fourth line:
\begin{equation}\label{calculforgamma}
\dcb
&&
\mathbb{E}[Ff(\boldsymbol{\tau}_I)\ind_{\{\tau_i\leq t<\tau_j, i\in I, j\in J\}}\ind_{\{s<\tau_{l}\}}]\\
&=&
\mathbb{E}[Ff(\boldsymbol{\tau}_I)\ind_{\{\tau_i\leq t, i\in I\}}\mathbb{E}[\ind_{\{t<\tau_j, j\in J\}}\ind_{\{s<\tau_{l}\}}|\mathcal{B}_t\vee\sigma(\boldsymbol{\tau}_I)]]\\
&=&
\mathbb{E}[Ff(\boldsymbol{\tau}_I)\ind_{\{\tau_i\leq t, i\in I\}}\Phi_{J, \rho^I,\sigma^I}(Z^{j,I}_t(u_{j}), j\in J)]\\
&&\mbox{where $u_{j}=t$ except $u_{l}=s$,}\\
&=&
\mathbb{E}[Ff(\boldsymbol{\tau}_I)\ind_{\{\tau_i\leq t<\tau_j, i\in I, j\in J\}}\frac{\Phi_{J, \rho^I,\sigma^I}(Z^{j,I}_t(u_{j}), j\in J)}{\Phi_{J, \rho^I,\sigma^I}(Z^{j,I}_t(t), j\in J)}]\\

&=&
\mathbb{E}[Ff(\boldsymbol{\tau}_I)\ind_{\{\tau_i\leq t<\tau_j, i\in I, j\in J\}}\frac{\Phi_{\mathcal{J}_{t}, \rho_{t},\sigma_{t}}(Z^{j,\mathcal{I}_{t}}_t(u_{j}), j\in \mathcal{J}_{t})}{\Phi_{\mathcal{J}_{t}, \rho_{t},\sigma_{t}}(Z^{j, \mathcal{I}_{t}}_t(t), j\in \mathcal{J}_{t})}].
\dce
\end{equation}
With the formula (\ref{Gtsplit}), we conclude\bel
&\mathbb{E}[\ind_{\{t<\tau_{l}\leq s\}}|\mathcal{G}_{t}]
=
\mathbb{E}[\ind_{\{t<\tau_{l}\}}|\mathcal{G}_{t}] - \mathbb{E}[\ind_{\{s<\tau_{l}\}}|\mathcal{G}_{t}]
\\&\qqq=
\ind_{\{t<\tau_{l}\}}(1 - \frac{\Phi_{\mathcal{J}_{t}, \rho_{t},\sigma_{t}}(Z^{j,\mathcal{I}_{t}}_t(u_{j}), j\in \mathcal{J}_{t})}{\Phi_{\mathcal{J}_{t}, \rho_{t},\sigma_{t}}(Z^{j, \mathcal{I}_{t}}_t(t), j\in \mathcal{J}_{t})}).
\eel
The stated result follows by an application of the Laplace formula 
of 
\citeN[Chapter V, Theorem T54]{Dellacherie72} (see also \citeN{dd:ce} or \citeN{kn:cc}).~\finproof

\subsection{Computation of the drift of the Brownian motion}

Next we study the processes $B^i, i\in N$, in the filtration $\mathbb{G}$. Thanks to Theorem \ref{l:explic}, the DGC model is a multivariate density model. According to \citeN{Jacod87}, this implies the following:

\bl
The processes $B^i, i\in N,$ are $\mathbb{G}$ semimartingales.~\finproof 
\el

By virtue of \citeN[Theorem 6.4]{JeanblancSong12}, another consequence of the multivariate density property is the martingale representation property.

\begin{theorem}
Let $W^i$, for $i\in N$, denote the martingale part in $\mathbb{G}$ of $B^i$. Let \index{m@$M^i$}$$
dM^i_t=d\ind_{{\tau_i\leq t}} -
 \gamma^i_t dt,\ \  t>0,
$$
where the process $\gamma^i$ is defined in (\ref{QoverSci}). Then, the martingale representation property holds in $\mathbb{G}$ with respect to $(W^i, M^i, i\in N)$.
\end{theorem}

This section is devoted to the computation of the martingales $W^i$. 
We begin with the following remark on the Gaussian processes $B^i$ (cf. Lemma \ref{l:Gauss}). 
%\b{(modulo consideration of brackets of local martingales here instead of correlations there)}:
\bl
For $J\subset N$ and $I=N\setminus J$, the family of processes 
\begin{equation}\label{Bindsplit}
\left(B^{j} - \frac{\varrho}{(|I|-1)\varrho + 1}\sum_{i\in I}B^{i}\right)_{j\in J}
\end{equation}
is a continuous L\'evy process
(multivariate Brownian motion with drift) independent of $\sigma(B^{i}, i\in I)$, of homogeneous marginal variances and pairwise correlations, equal to,
respectively,
\begin{equation}
(\sigma^I)^2= (1 - \varrho)\frac{|I|\varrho +1}{|I|\varrho +1 - \varrho}  \mbox{ and }\ \
\rho^I= \frac{\varrho}{|I|\varrho +1}.~\finproof
\end{equation} 
\el
\proof This
follows by computing the brackets of the continuous local martingales and applying the L\'evy processes characterization.~\finproof
\bl 
For $k\in I$ and $0\le t\le s\le s'$, 
$$
\dcb
&&\mathbb{E}[B^k_{s'} - B^k_{s}|\mathcal{B}_t\vee\sigma(\boldsymbol{\tau}_I)]
=
(\frac{1}{\alpha^2_{t}}\int_{s}^{s'}\!\!\!\varsigma_{u}du)\ \overline{m}^k_{t}.
\dce
$$
\el
%\proof
%\b{Recall that $$
%\lambda^I = \frac{\varrho}{(|I|-1)\varrho + 1},\
%(\sigma^I)^2=(1 - \varrho)\frac{|I|\varrho +1}{|I|\varrho +1 - \varrho},\
%\rho^I= \frac{\varrho}{|I|\varrho +1}.
%$$
\proof  
For $k\in I$, for $0\leq t\leq s\leq s'<\infty$, 
\begin{equation}\label{Bprojmbar}
B^k_{s'} - B^k_{s} - (\frac{1}{\alpha^2_{t}}\int_{s}^{s'}\!\!\!\varsigma_{u}du)\ \overline{m}^k_{t} 
\end{equation}
is a centered Gaussian random variable, independent of $\overline{m}^k_{t}$, with variance$$
(s' - s) - 2 \frac{1}{\alpha^2_{t}}(\int_{s}^{s'}\!\!\!\varsigma_{u}du)^2
+
\frac{1}{\alpha^2_{t}}(\int_{s}^{s'}\!\!\!\varsigma_{u}du)^2
=
(s' - s) - \frac{1}{\alpha^2_{t}}(\int_{s}^{s'}\!\!\!\varsigma_{u}du)^2.
$$
Hence, for $k\in I$, 
$$
\dcb
&&\mathbb{E}[B^k_{s'} - B^k_{s}|\mathcal{B}_t\vee\sigma(\boldsymbol{\tau}_I)]
=
\mathbb{E}[B^k_{s'} - B^k_{s}|\mathcal{B}_t\vee\sigma(\overline{m}^i_{t}, i\in I)]\\
&=&
\mathbb{E}[B^k_{s'} - B^k_{s}|\mathcal{B}_t\vee\sigma(\overline{m}^k_{t})\vee \sigma(\xi_{i}, i\in I\setminus \{k\})]\\
&&\mbox{where $\xi_{i}$ form a Gaussian family independent of $\mathcal{B}_t\vee\sigma(B^k)$, constructed with (\ref{Bindsplit}),}\\
&=&
\mathbb{E}[B^k_{s'} - B^k_{s}|\sigma(\overline{m}^k_{t})]
=
(\frac{1}{\alpha^2_{t}}\int_{s}^{s'}\!\!\!\varsigma_{u}du)\ \overline{m}^k_{t}\
\mbox{because of (\ref{Bprojmbar})}.~\finproof
\dce
$$

In the sequel we find it sometimes convenient 
to denote stochastic integration (or integration against measures) by
${\centerdot}$ and the Lebesgue measure on the half-line by $\boldsymbol{\lambda}$.

\begin{theorem}\label{lem:fbt} 
For $J\subset N$ and $k\in J$, define the function
$$
\mathfrak{b}^k_{J}(\mathbf{z}, x)
\eq 
\mathbb{E}[\ind_{\{z_{j}<\xi_{j}, j\in J\}}(\xi_{k} + x) ]
\sp
x\in\mathbb{R},\,\mathbf{z}=(z_{j}, j\in J) ,
$$
where $(\xi_{j}, j\in J)$ is a Gaussian family of homogeneous marginal variances $(\sigma^I)^2$ and pairwise correlations $\rho^I$. 
For any $k\in N$, define the process 
$$\beta^k_{t}
\eq 
\frac{\varsigma(t)}{\alpha(t)}
\left(\ind_{\{k\in \mathcal{I}_{t}\}}\ \frac{\overline{m}^k_{t}}{\alpha(t)}
+
\ind_{\{k\notin \mathcal{I}_{t}\}}\frac{\mathfrak{b}^k_{\mathcal{J}_{t}}((Z^{j,\mathcal{I}_{t}}_{t}(t), j\in \mathcal{J}_{t}),\ \lambda^{\mathcal{I}_{t}} \sum_{i\in \mathcal{I}_{t}}\frac{\overline{m}^i_{t}}{\alpha(t)})}
{\Phi_{\mathcal{J}_{t}, \rho_{t},\sigma_{t}}(Z^{j, \mathcal{I}_{t}}_t(t), j\in \mathcal{J}_{t})}
\right)\sp t\in\R_+.
$$
Then, $W^k = B^k - \beta^{k}{\centerdot}\boldsymbol{\lambda}$.
\end{theorem}

\proof 
For $0\leq t\leq s\leq s'<\infty$, for any bounded $\mathcal{B}_t$ measurable function $F$ and measurable bounded function $f$, we compute
$$
\dcb
&&
\mathbb{E}[Ff(\boldsymbol{\tau}_I)\ind_{\{\tau_i\leq t<\tau_j, i\in I, j\in J\}}(B^k_{s'} - B^k_{s})]\\
&=&
\mathbb{E}[Ff(\boldsymbol{\tau}_I)\ind_{\{\tau_i\leq t, i\in I\}}\ind_{\{t<\tau_j, j\in J\}}\mathbb{E}[(B^k_{s'} - B^k_{s}) |\mathcal{B}_t\vee\sigma(\boldsymbol{\tau}_N)]]\\

&=&
\mathbb{E}[Ff(\boldsymbol{\tau}_I)\ind_{\{\tau_i\leq t, i\in I\}}\ind_{\{t<\tau_j, j\in J\}}(\frac{1}{\alpha^2_{t}}\int_{s}^{s'}\!\!\!\varsigma_{u}du)\ \overline{m}^k_{t}].
\dce
$$
If $k\in I$, $\overline{m}^k_{t}\in \mathcal{B}_t\vee\sigma(\boldsymbol{\tau}_I)$. If $k\notin I$,
$$
\dcb
&&
\mathbb{E}[Ff(\boldsymbol{\tau}_I)\ind_{\{\tau_i\leq t, i\in I\}}\ind_{\{t<\tau_j, j\in J\}}(\frac{1}{\alpha^2_{t}}\int_{s}^{s'}\!\!\!\varsigma_{u}du)\ \overline{m}^k_{t}]\\

&=&
\mathbb{E}[Ff(\boldsymbol{\tau}_I)\ind_{\{\tau_i\leq t, i\in I\}}(\frac{1}{\alpha(t)}\int_{s}^{s'}\!\!\!\varsigma_{u}du)\ \mathbb{E}[\ind_{\{t<\tau_j, j\in J\}}\frac{\overline{m}^k_{t}}{\alpha(t)} |\mathcal{B}_t\vee\sigma(\boldsymbol{\tau}_I)]]\\

&=&
\mathbb{E}[Ff(\boldsymbol{\tau}_I)\ind_{\{\tau_i\leq t, i\in I\}}(\frac{1}{\alpha(t)}\int_{s}^{s'}\!\!\!\varsigma_{u}du)\\ 
&&
\mathbb{E}[\ind_{\{Z^{j,I}_{t}(t)<\frac{\overline{m}^j_{t}}{\alpha(t)} - \lambda^I \sum_{i\in I}\frac{\overline{m}^i_{t}}{\alpha(t)}, j\in J\}}(\frac{\overline{m}^k_{t}}{\alpha(t)} - \lambda^I \sum_{i\in I}\frac{\overline{m}^i_{t}}{\alpha(t)} + \lambda^I \sum_{i\in I}\frac{\overline{m}^i_{t}}{\alpha(t)}) |\mathcal{B}_t\vee\sigma(\boldsymbol{\tau}_I)]]\\

&=&
\mathbb{E}[Ff(\boldsymbol{\tau}_I)\ind_{\{\tau_i\leq t, i\in I\}}(\frac{1}{\alpha(t)}\int_{s}^{s'}\!\!\!\varsigma_{u}du)\
\mathbb{E}[\ind_{\{Z^{j,I}_{t}(t)<\xi_{j}, j\in J\}}(\xi_{k} + \lambda^I \sum_{i\in I}\frac{\overline{m}^i_{t}}{\alpha(t)}) ]]\\

&=&
\mathbb{E}[Ff(\boldsymbol{\tau}_I)\ind_{\{\tau_i\leq t, i\in I\}}(\frac{1}{\alpha(t)}\int_{s}^{s'}\!\!\!\varsigma_{u}du)\
\mathfrak{b}^k_{J}((Z^{j,I}_{t}(t), j\in J),\ \lambda^I \sum_{i\in I}\frac{\overline{m}^i_{t}}{\alpha(t)}) ]\\

&=&
\mathbb{E}[Ff(\boldsymbol{\tau}_I)\ind_{\{\tau_i\leq t, i\in I\}}\ind_{\{t<\tau_j, j\in J\}}(\frac{1}{\alpha(t)}\int_{s}^{s'}\!\!\!\varsigma_{u}du)\
\frac{\mathfrak{b}^k_{J}((Z^{j,I}_{t}(t), j\in J),\ \lambda^I \sum_{i\in I}\frac{\overline{m}^i_{t}}{\alpha(t)})}
{\Phi_{J, \rho^I,\sigma^I}(Z^{j, I}_t(t), j\in J)}
]\\

\dce
$$
This, combined with the formula (\ref{Gtsplit}), implies
\bel&
\mathbb{E}[(B^k_{s'} - B^k_{s})|\mathcal{G}_{t}]=(\frac{1}{\alpha(t)}\int_{s}^{s'}\!\!\!\varsigma_{u}du)\times
\\&
\left(\ind_{\{k\in \mathcal{I}_{t}\}}\ \frac{\overline{m}^k_{t}}{\alpha(t)}
+
\ind_{\{k\notin \mathcal{I}_{t}\}}\frac{\mathfrak{b}^k_{\mathcal{J}_{t}}((Z^{j,\mathcal{I}_{t}}_{t}(t), j\in \mathcal{J}_{t}),\ \lambda^{\mathcal{I}_{t}} \sum_{i\in \mathcal{I}_{t}}\frac{\overline{m}^i_{t}}{\alpha(t)})}
{\Phi_{\mathcal{J}_{t}, \rho_{t},\sigma_{t}}(Z^{j, \mathcal{I}_{t}}_t(t), j\in \mathcal{J}_{t})}
\right).\eel

The $\mathbb{G}$ drift of $B^k$ is obtained as the differential 
of the above 
with respect to Lebesgue measure, i.e.
$$
\frac{\varsigma(t)}{\alpha(t)}
\left(\ind_{\{k\in \mathcal{I}_{t}\}}\ \frac{\overline{m}^k_{t}}{\alpha(t)}
+
\ind_{\{k\notin \mathcal{I}_{t}\}}\frac{\mathfrak{b}^k_{\mathcal{J}_{t}}((Z^{j,\mathcal{I}_{t}}_{t}(t), j\in \mathcal{J}_{t}),\ \lambda^{\mathcal{I}_{t}} \sum_{i\in \mathcal{I}_{t}}\frac{\overline{m}^i_{t}}{\alpha(t)})}
{\Phi_{\mathcal{J}_{t}, \rho_{t},\sigma_{t}}(Z^{j, \mathcal{I}_{t}}_t(t), j\in \mathcal{J}_{t})}
\right)dt.~\finproof
$$
 
\section{Reduced DGC Model}

We now study the invariance properties of the DGC model. In this perspective, the market information before the default event of the bank or of its counterparty is modeled by the filtration  $\ff=(\mathcal{F}_t)_{t\geq 0}$, where 
\beql{e:dgcff}
\mathcal{F}_{t}=\cap_{s>t}\big(\mathcal{B}_{s} \vee \bigvee_{i\in \N^\star } \big( \sigma(\tau_i
 \wedge s)\big),
\eeql
augmented so as to satisfy the usual conditions. 

Because of the multivariate density property of the family of $(\tau^j, j\in N^\star)$ with respect to the filtration $\mathbb{B}$ (same proof as Theorem \ref{l:explic}), the computations we have made in $\mathbb{G}$ in the previous section can be made similarly in $\mathbb{F}$. In particular, the following splitting formula holds (cf.~(\ref{Gtsplit})): for any $t>0$ and $I\subset N^\star,$ writing $J= N^\star \setminus I$, 
\begin{equation}\label{Ftsplit}
\dcb
&&\{\tau_i\leq t<\tau_j: i\in I, j\in J\}\cap\mathcal{F}_t=
\{\tau_i\leq t<\tau_j: i\in I, j\in J\}\cap\mathcal{B}_t\vee\sigma(\boldsymbol{\tau}_I).
\dce
\end{equation}
Moreover, the so-called condition (H') holds, i.e.
the processes $B^k, k\in N,$ are $\mathbb{F}$ semimartingales, and the random times $\tau_{j}, j\in N^\star,$ are $\mathbb{F}$ totally inaccessible stopping times, as stated in the following lemma. For $t>0$, let$$
\dcb
\mathcal{I}^\star_{t} = \{i\in N^\star: \tau_{i}\leq t\},\ \mathcal{J}^\star_{t} = N^\star \setminus \mathcal{I}^\star_{t},\
\rho^\star_t = {\rho}^{\mathcal{I}^\star_{t}},\
\sigma^\star_t = \sigma^{\mathcal{I}^\star_{t}}.
\dce
$$

\bl\label{BtauinF} 
For any $k\in N$, the process $\overline{W}^k_{t} = B^k_{t} - \int_{0}^t \overline{\beta}^{k}_{s}ds, \ t\geq 0$ is an $\mathbb{F}$ local martingale, where 
\begin{eqnarray*} 
\overline{\beta}^{k}_{t}
\eq 
\frac{\varsigma(t)}{\alpha(t)}
\left(\ind_{\{k\in \mathcal{I}^\star_{t}\}}\ \frac{\overline{m}^k_{t}}{\alpha(t)}
+
\ind_{\{k\notin \mathcal{I}^\star_{t}\}}\frac{\mathfrak{b}^k_{\mathcal{J}^\star_{t}}((Z^{j,\mathcal{I}^\star_{t}}_{t}(t), j\in \mathcal{J}^\star_{t}),\ \lambda^{\mathcal{I}^\star_{t}} \sum_{i\in \mathcal{I}^\star_{t}}\frac{\overline{m}^i_{t}}{\alpha(t)})}
{\Phi_{\mathcal{J}^\star_{t}, \rho^\star_{t},\sigma^\star_{t}}(Z^{j, \mathcal{I}^\star_{t}}_t(t), j\in \mathcal{J}^\star_{t})}
\right)\sp t\in\R_+.
\end{eqnarray*}
For $j\in N^\star$, $\tau_{j}$ is an $\mathbb{F}$ totally inaccessible stopping time and the process \index{m@$M^i$}$
d\overline{M}^j_t=d\ind_{{\tau_j\leq t}} -
 \overline{\gamma}^j_t dt,\ \  t>0,
$
is an $\mathbb{F}$ local martingale, where 
\begin{eqnarray}
 \label{overlinegamma} 
&&
\overline{\gamma}^j_{t}
\eq 
\ind_{\{t<\tau_j \}}
 \frac{\dot{h} _j(t)}{\alpha(t)} 
 \dl_{\mathcal{J}^\star_{t},\rho^\star_t,\sigma^\star_t}\big( Z^{j, \mathcal{I}^\star_{t}}_t(t), j\in \mathcal{J}^\star_{t} \big)\sp t\in\R_+.
\end{eqnarray} 
The family of processes $\overline{W}^k, k\in N$ and $\overline{M}^j, j\in N^\star,$ has the martingale representation property in the filtration $\mathbb{F}$.~\finproof
\el

\subsection{The Az\'ema supermartingale}

Our next aim is to compute the Az\'ema supermartingale of the random time $\tau_{-1}\wedge \tau_{0}$ in the filtration $\mathbb{F}$, i.e., 
$$
\mathbb{E}[\ind_{\{t<\tau_{-1}\wedge \tau_{0}\}}|\mathcal{F}_{t}], \ t\geq 0.
$$
%The following computation is carried out in the same way as (\ref{calculforgamma}). 

\bl
The Az\'ema supermartingale of the random time $\tau_{-1}\wedge \tau_{0}$ in the filtration $\mathbb{F}$ is given by
\begin{equation}\label{expressionS}
S_{t}
\eq 
\frac{\Phi_{\mathcal{J}^\star_{t}\cup\{-1,0\}, \rho^\star_{t},\sigma^\star_{t}}(Z^{j,\mathcal{I}^\star_{t}}_t(t), j\in \mathcal{J}^\star_{t}\cup\{-1,0\})}{\Phi_{\mathcal{J}^\star_{t}, \rho^\star_{t},\sigma^\star_{t}}(Z^{j, \mathcal{I}^\star_{t}}_t(t), j\in \mathcal{J}^\star_{t})}, \ t\geq 0.
\end{equation}
In particular, the Az\'ema supermartingale $S$ is positive.
\el

\proof
For any bounded $\mathcal{B}_t$ measurable functions $F$ and measurable bounded function $f$, we compute (cf. (\ref{calculforgamma}))
$$
\dcb
&&
\mathbb{E}[Ff(\boldsymbol{\tau}_{I})\ind_{\{\tau_i\leq t<\tau_j, i\in I, j\in J\}}\ind_{\{t<\tau_{-1}\wedge \tau_{0}\}}]\\
&=&
\mathbb{E}[Ff(\boldsymbol{\tau}_{I})\ind_{\{\tau_i\leq t: i\in I\}}\mathbb{E}[\ind_{\{t<\tau_j: j\in J\}} \ind_{\{t<\tau_{-1}\wedge \tau_{0}\}} |\mathcal{B}_t\vee\sigma(\boldsymbol{\tau}_I)]]\\
&=&
\mathbb{E}[Ff(\boldsymbol{\tau}_{I})\ind_{\{\tau_i\leq t: i\in I\}}\Phi_{J\cup\{-1,0\}, \rho^{I},\sigma^{I}}(Z^{j,I}_t(t): j\in J\cup\{-1,0\})]\\

&=&
\mathbb{E}[Ff(\boldsymbol{\tau}_{I})\ind_{\{\tau_i\leq t<\tau_j: i\in I, j\in J\}}\frac{\Phi_{J\cup\{-1,0\}, \rho^{I},\sigma^{I}}(Z^{j,I}_t(t): j\in J\cup\{-1,0\})}{\Phi_{J, \rho^{I},\sigma^{I}}(Z^{j,I}_t(t): j\in J)}]\\

&=&
\mathbb{E}[Ff(\boldsymbol{\tau}_I)\ind_{\{\tau_i\leq t<\tau_j: i\in I, j\in J\}}\frac{\Phi_{\mathcal{J}^\star_{t}\cup\{-1,0\}, \rho^\star_{t},\sigma^\star_{t}}(Z^{j,\mathcal{I}^\star_{t}}_t(t): j\in \mathcal{J}^\star_{t}\cup\{-1,0\})}{\Phi_{\mathcal{J}^\star_{t}, \rho^\star_{t},\sigma^\star_{t}}(Z^{j, \mathcal{I}^\star_{t}}_t(t): j\in \mathcal{J}^\star_{t})}],
\dce
$$
where conditioning and the tower rule are used in the next-to-last identity.
With the formula (\ref{Ftsplit}), we conclude$$
\mathbb{E}[\ind_{\{t<\tau_{-1}\wedge \tau_{0}\}}|\mathcal{F}_{t}]
=
\frac{\Phi_{\mathcal{J}^\star_{t}\cup\{-1,0\}, \rho^\star_{t},\sigma^\star_{t}}(Z^{j,\mathcal{I}^\star_{t}}_t(t): j\in \mathcal{J}^\star_{t}\cup\{-1,0\})}{\Phi_{\mathcal{J}^\star_{t}, \rho^\star_{t},\sigma^\star_{t}}(Z^{j, \mathcal{I}^\star_{t}}_t(t): j\in \mathcal{J}^\star_{t})}.~\finproof
$$

Let $\nu\eq \frac{1}{S}\centerdot{S^c}$, where $S^c$ denotes
the continuous martingale component of the $(\ff,\Q)$ Az\'ema supermartingale $S$.

\bl\label{l:explicred} 
We have
$$\label{QoverSbsuite}
\dcb
d\nu_t
&=&
\sum_{j \in\mathcal{J}^\star_{t-}\cup\{-1,0\}} \psi^j_{\mathcal{J}^\star_{t-}\cup\{-1,0\},\rho^\star_t,\sigma^\star_t}\big(Z^{j,\mathcal{I}^\star_{t-}}_t(t): j\in \mathcal{J}^\star_{t}\cup\{-1,0\}\big)
d\Zm^{j, \mathcal{I}^\star_{t}}_t\\
&&-
\sum_{j \in\mathcal{J}^\star_{t-}} \psi^j_{\mathcal{J}^\star_{t-},\rho^\star_t,\sigma^\star_t}\big(Z^{j,\mathcal{I}^\star_{t-}}_t(t): j\in \mathcal{J}^\star_{t}\big)
d\Zm^{j, \mathcal{I}^\star_{t}}_t,\\
\dce
$$
where, for $I\subset N^\star$, $\Zm^{j,I}_t$ denotes the martingale part of 
$\left(-\frac{1}{\mynu(t)}dm^j_t + \frac{\varrho}{(|I|-1)\varrho+1}
\sum_{i\in I} \frac{1}{\mynu(t)}dm^i_t\right)$ in $\mathbb{F}$.
\el
 
\proof 
To obtain $dS^c_{t}$ (which is then divided by $S_t$), it suffices to apply It\^o calculus to the expression \qr{expressionS} of $S$ on every random interval where $\mathcal{I}^\star_{t-}$ is constant. Note that, knowing $t$ is in such an interval, $\boldsymbol{\tau}_{\mathcal{I}^\star_{t-}}$ is in $\mathcal{F}_{t}$. Also note that the jumps of $S_{t}$ triggered by the jumps of  $\mathcal{I}^\star_{t-}$ have no impact here, because $S^c$ is a continuous local martingale.~\finproof

\subsection{$\mathbb{F}$ reductions of $\beta^k, \gamma^j$}\label{s:a1}

\bl
The triplet $(\tau_{-1}\wedge \tau_{0},\mathbb{F}, \mathbb{G})$ satisfies the condition (B). 
\el

\proof To check the condition (B), by the monotone class theorem, we only need consider the elementary $\gg$ predictable processes of the form
$U = \nu f(\tau_{-1}\wedge s, \tau_0 \wedge s)\ind_{(s,t]},$
for an $\F_s$ measurable random variable $F $ and a Borel function $f$. Since
$U\ind_{(0,\tau]} =  {F } f(s, s)\ind_{(s,t]} \ind_{(0,\tau]},$
we may take $U^{\prime}= {F } f(s, s)\ind_{(s,t]}$ in the condition (B).~\finproof\\

Next we consider the reduction of the processes $\beta^k, \gamma^j$ in the filtration $\mathbb{F}$. Notice that, for $t<\tau_{-1}\wedge \tau_{0}$,$$
\mathcal{I}_{t}=\mathcal{I}^\star_{t}, \ \mathcal{J}_{t}=\mathcal{J}^\star_{t}\cup\{-1,0\}.
$$
Therefore, the following lemma holds.

\bl\label{Freductions}
The $\mathbb{F}$ reduction of $\gamma^j, j\in N^\star,$ is
\begin{eqnarray}\label{expressiontildegamma}
\widetilde{\gamma}^j_{t}
\eq 
\ind_{\{t<\tau_j \}}
 \frac{\dot{h} _j(t)}{\alpha(t)} 
 \psi^j_{\mathcal{J}^\star_{t}\cup\{-1,0\},\rho^\star_t,\sigma^\star_t}\big( Z^{j, \mathcal{I}^\star_{t}}_t(t), j\in \mathcal{J}^\star_{t}\cup\{-1,0\} \big)\sp t\in\R_+.
\end{eqnarray}
Similarly, the $\mathbb{F}$ reduction of $\beta^k, k\in N,$ is\bel
&\widetilde{\beta}^{k}_{t}
\eq 
\frac{\varsigma(t)}{\alpha(t)}\times\left(\ind_{\{k\in \mathcal{I}^\star_{t}\}}\ \frac{\overline{m}^k_{t}}{\alpha(t)}
+\right.\\&\qqq\left.
\ind_{\{k\notin \mathcal{I}^\star_{t}\}}\frac{\mathfrak{b}^k_{\mathcal{J}^\star_{t}\cup\{-1,0\}}((Z^{j,\mathcal{I}^\star_{t}}_{t}(t), j\in \mathcal{J}^\star_{t}\cup\{-1,0\}),\ \lambda^{\mathcal{I}^\star_{t}} \sum_{i\in \mathcal{I}^\star_{t}}\frac{\overline{m}^i_{t}}{\alpha(t)})}
{\Phi_{\mathcal{J}^\star_{t}\cup\{-1,0\}, \rho^\star_{t},\sigma^\star_{t}}(Z^{j, \mathcal{I}^\star_{t}}_t(t), j\in \mathcal{J}^\star_{t}\cup\{-1,0\})}
\right)\sp t\in\R_+.~\finproof
\eel
\el
Note that the processes ${\gamma}^j, \widetilde{\gamma}^j$ and $\beta^k, \widetilde{\beta}^k$ are c\`adl\`ag. The next result shows that the process $\beta^k$ (and consequently $\widetilde{\beta}$) is linked with $\overline{\beta}^k$ through the process $\nu$.

\bl\label{beta-beta}
For $k\in N$, $$
\int_{0}^t \widetilde{\beta}^k_{s} ds
=
\int_{0}^t \overline{\beta}^k_{s} ds + \cro{B^k, \nu}_{t}, \ t\in [0,\tau_{-1}\wedge \tau_{0}].
$$
\el

\proof
Notice that $B^k$ is a continuous process. By the Jeulin--Yor formula {(see e.g.~\citeN[no 77 Remarques b)]{DellacherieMaisonneuveMeyer92}),}$$
B^k_{t} - \int_{0}^t \overline{\beta}^k_{s} ds - \cro{B^k, \nu}_{t}, \ t\in [0,\tau_{-1}\wedge \tau_{0}],
$$
defines a $\mathbb{G}$ local martingale. But, acccording to Theorem \ref{lem:fbt}, the drift of $B^k$ in $\mathbb{G}$ is $\int_{0}^t {\beta}^k_{s} ds, t\geq 0$. We conclude that $$
\int_{0}^t \widetilde{\beta}^k_{s} ds
=
\int_{0}^t {\beta}^k_{s} ds
=
\int_{0}^t \overline{\beta}^k_{s} ds + \cro{B^k, \nu}_{t}
$$
for $t\in [0,\tau_{-1}\wedge \tau_{0}]$.
\finproof\\

Knowing the $\mathbb{F}$ reductions $\widetilde{\beta}^{k}$ and $\widetilde{\gamma}^{j}$ of $\beta^k$ and $\gamma^j$, 
in view  of the martingale representation property in $\mathbb{F},$
accounting also for the avoidance of $\tau_{-1}\wedge \tau_{0}$ and $\tau_{j}, j\in N^\star$, the strategy for constructing an invariance probability measure $\P$ becomes clear. It is enough to find a probability measure $\P$ equivalent to $\Q$ on $\mathcal{F}_{T}$ (given a constant $T>0$) such that the $(\mathbb{F},\mathbb{P})$ drift of $B^k, k\in N,$ is $\widetilde{\beta}^{k}$ and the $(\mathbb{F},\mathbb{P})$ compensator of $\tau^j, j\in N^\star$, has the density process $\widetilde{\gamma}^{j}$.

To implement this idea, the following estimates will be useful.

\bl\label{corB} 
There exists a constant $C>0$ 
such that
\beqa 
&&\cro{\nu}_t 
\leq C ( \sum_{i\in N}\sup_{0<s\leq t}|m^{i}_{s}|^2+1)t \label{l:cro}\eeqa
and for $0\leq r\leq t$ and $j\in N^\star$
\beqa\label{l:lbds}
&&\widetilde{\gamma}^j_r\vee \bar{\gamma}^j_r
 \leq C(\sum_{i\in N}\sup_{0<s\leq t}|m^{i}_{s}|+1)  ,
 \\&& \nonumber {\widetilde{\gamma}^j_r
\ln(\widetilde{\gamma}^j_r \vee \overline{\gamma}^j_r) 
\leq C\sum_{i\in N}\sup_{0<s\leq t}(|m^{i}_{s}|+1)\ln(|m^{i}_{s}|+1) }.
\eeqa
\el

\proof Applying Lemma \ref{estim} to the formula \qr{QoverSbsuite} and noting that the function $\mynu,$ continuous and positive, is bounded away from 0 on $[0,T],$
we obtain, for positive constants $C$ that may change from place to place,
\bel
&\cro{\nu}_t 
 %=\int_0^t\frac{1}{2S_s^2} d\cro{S^c}_s
 \leq 
C \int_0^t(\sum_{I\subseteq N}\sum_{j\in N\setminus I}|Z^{j,I}_{s}(s)|+1)^2ds \leq C (\sum_{I\subseteq N}\sum_{j\in N\setminus I}\sup_{0<s\leq t}|Z^{j,I}_{s}(s)|+1)^2t,
\eel
which yields \qr{l:cro}.
Applying Lemma \ref{estim} 
to the formulas (\ref{overlinegamma}) for $\overline{\gamma}^j$ and   (\ref{expressiontildegamma}) for $\tilde{\gamma}^j$, we obtain
the first line in \qr{l:lbds}),
whence the second line follows from
\bel&
\widetilde{\gamma}^j_r
\ln(\widetilde{\gamma}^j_r \vee \overline{\gamma}^j_r)\leq 
 C(\max_{i\in N}\sup_{0<s\leq t}|m^{i}_{s}|+1) 
\ln \big( C(\max_{i\in N}\sup_{0<s\leq t}|m^{i}_{s}|+1) \big)\\&\qqq
=\max_{i\in N}\sup_{0<s\leq t}
C(|m^{i}_{s}|+1)\ln(C|m^{i}_{s}|+1).~\ok 
\eel

Notice that the processes $\overline{\gamma}^j$ are positive. Consider the $\mathbb{F}$ local martingale $\mu\eq \nu+\sum_{j\in N^\star}
(\frac{\tilde{\gamma}^j_{-}}{\overline{\gamma}^j_{-}}-1)\centerdot \overline{M}^j$. 
%As a consequence of Lemma \ref{corB}, we have:

\bl\label{tt1}
The Dol\'eans-Dade exponential $\mathcal{E}(\mu)$ is a true $(\mathbb{F},\mathbb{Q})$ martingale. 
\el

\proof
Following \citeN[Theorem III.1]{LepingleMemin78}, we consider $$
\dcb
\sum_{j\in N^\star}
\left(
(1+(\frac{\tilde{\gamma}^j_{\tau_{j}-}}{\overline{\gamma}^j_{\tau_{j}-}}-1))
\ln(1+(\frac{\tilde{\gamma}^j_{\tau_{j}-}}{\overline{\gamma}^j_{\tau_{j}-}}-1)) 
- (\frac{\tilde{\gamma}^j_{\tau_{j}-}}{\overline{\gamma}^j_{\tau_{j}-}}-1)
\right)
\ind_{\{\tau_{j}\leq t\}},
\dce
$$
and its $\mathbb{F}$ predictable dual projection$$
\dcb
M_{t}
&\eq &
\sum_{j\in N^\star}
\int_{0}^t
\left(
(1+(\frac{\tilde{\gamma}^j_{s-}}{\overline{\gamma}^j_{s-}}-1))
\ln(1+(\frac{\tilde{\gamma}^j_{s-}}{\overline{\gamma}^j_{s-}}-1)) 
- (\frac{\tilde{\gamma}^j_{\tau_{j}-}}{\overline{\gamma}^j_{s-}}-1)
\right)
\overline{\gamma}^j_{s-}ds\\

&=&
\sum_{j\in N^\star}
\int_{0}^t
\left(
{\tilde{\gamma}^j_{s-}}
(\ln({\tilde{\gamma}^j_{s-}})
-
\ln({\overline{\gamma}^j_{s-}}) )
- ({\tilde{\gamma}^j_{\tau_{j}-}}-{\overline{\gamma}^j_{s-}})
\right)
ds.

\dce
$$
Combining Lemma \ref{corB} and Lemma \ref{bab}, we prove that $e^{\frac{1}{2}\cro{\nu}_{t}+M_{t}}$ is $\mathbb{Q}$ integrable for a sufficiently small $t=t_{0}>0$. According to \citeN[Theorem III.1]{LepingleMemin78}, we conclude that $\mathbb{E}[\mathcal{E}(\mu)_{t_{0}}]=1$. The same argument applied with the conditional probability $\mathbb{Q}[\cdot |\mathcal{F}_{t_{0}}]$ instead of $\mathbb{Q}$ proves that $\mathbb{E}[\mathcal{E}(\mu)_{2t_{0}} |\mathcal{F}_{t_{0}}]=1$. Iterating, we arrive at $\mathbb{E}[\mathcal{E}(\mu)_{kt_{0}}]=1$ for any integer $k>0$.
\finproof

\subsection{The invariance probability measure}\label{s:a}

We have proved that $\mathcal{E}(\mu)$ is an $(\mathbb{F},\mathbb{Q})$ true martingale. We can then define a new probability measure $\mathbb{P}\eq \mathcal{E}(\mu).\mathbb{Q}$ on $\cF_T$.

%Notice that, as demonstrated in \citeN{Emery80}, a stochastic integral in the sense of semimartingales with respect to a local martingale need not be a local martingale. \b{This justifies the distinction}, in the proof of the next result, between the stochastic integral in the sense of semimartingales and the stochastic integral in the sense of local martingales. See also \citeN{DelbaenSchachermayer}.
%Yan: même s'il définit la générale, dans les calculs most often en 2 bouts (afin précisément d'éviter le pbme) , contre FV et contre loc vol
\begin{theorem}\label{DGC2}
The probability measure $\mathbb{P}$ is an invariance probability measure for the DGC model $(\tau_{-1}\wedge \tau_{0},\mathbb{F}, \mathbb{G}, \mathbb{Q})$ on the horizon $[0,T],$ for any constant $T>0$.
\end{theorem} 

\proof 
By the Girsanov theorem, the intensity of $\tau_i$, $i\in N^{\star},$ in $\mathbb{F}$ under $\mathbb{P}$ is $\tilde{\gamma}^i$, while
the drift of $B^k, k\in N,$ in $\mathbb{F}$ under $\mathbb{P}$ is $(\overline{\themu}^k \centerdot \boldsymbol\lambda+ \cro{B^k,\nu})$.

Given a constant $T>0$, let us prove that the probability measure  $\mathbb{P}$ such that $\frac{d\mathbb{P}}{d\mathbb{Q}}=\cal E(\mu)$ is an invariance probability measure for the quadruplet $(\tau_{-1}\wedge \tau_{0},\mathbb{F}, \mathbb{G}, \mathbb{Q})$. According to \citeN[Corollary C.1]{CrepeySong15c}, we only need to consider the locally bounded $(\mathbb{F},\mathbb{P})$ local martingales $P$ in the condition (A). We write $$
\hat{W}^k=B^k-\overline{\themu}^k \centerdot \boldsymbol\lambda- \cro{B^k,\nu}\,\ k\in N, \mbox{ and }\
\hat{M}^j=\ind_{[\tau^j, \infty)}-\widetilde{\gamma}^j \centerdot \boldsymbol\lambda,\
k\in N,\ j\in N^\star.
$$
Thanks to Lemma \ref{beta-beta},
the stopped process $$
(\hat{W}^k)^{\tau_{-1}\wedge\tau_{0}-}
=
(B^k-{\themu}^k \centerdot \boldsymbol\lambda)^{\tau_{-1}\wedge\tau_{0}}
=
(W^k)^{\tau_{-1}\wedge\tau_{0}} 
$$ 
is a $(\mathbb{G},\mathbb{Q})$ local martingale. The $(\mathbb{G},\mathbb{Q})$ local martingale property of $$
(\hat{M}^j)^{\tau_{-1}\wedge\tau_{0}-}
=
(\ind_{[\tau^j, \infty)}-{\gamma}^j \centerdot \boldsymbol\lambda)^{\tau_{-1}\wedge\tau_{0}}
=
(M^j)^{\tau_{-1}\wedge\tau_{0}}
$$ 
(cf. Lemma \ref{Freductions}) is clear. 

As we did in Lemma \ref{BtauinF} under the probability $\mathbb{Q}$, it can be proven that the family of processes $
\hat{W}^k,$ $k\in N,$ and $\hat{M}^j,$ $j\in N^\star,$
has the martingale representation property in the filtration $\mathbb{F}$ under $\mathbb{P}$. Hence any $(\mathbb{F},\mathbb{P})$ local martingale $P$ is an stochastic integral in $\mathbb{F}$ of the processes $\hat{W}^k$ and of $\hat{M}^j$ under the probability measure $\mathbb{P}$. The natural idea is to say, then, $P^{\tau_{-1}\wedge\tau_{0}-}$ is the stochastic integral
% in the sense of local martingales 
in $\mathbb{G}$ of the processes $(W^k)^{\tau_{-1}\wedge\tau_{0}}$ and of $(M^j)^{\tau_{-1}\wedge\tau_{0}}$ under the probability $\mathbb{Q}$, so that $P^{\tau_{-1}\wedge\tau_{0}-}$ itself is a $(\mathbb{G},\mathbb{Q})$ local martingale. However, knowing the discussion in \citeN{JeulinYor79} about ``{\it faux amis}'' regarding enlargement of filtration and stochastic integrals, we have to be careful.
More precisely, we need to distinguish between the stochastic integral in the sense of semimartingales and the stochastic integral in the sense of local martingales, recalling from \citeN{Emery80} (cf. also \citeN[Theorem 2.9]{DelbaenSchachermayer}) that a stochastic integral in the sense of semimartingales with respect to a local martingale need not be a local martingale.

We can argue as follows. 
%Let us denote by $\hat{W}^k, k\in N$, the process $B^k-\overline{\themu}^k \centerdot \boldsymbol\lambda- \cro{B^k,\nu}$, and by $\hat{M}^j, j\in N^\star$, the process $\ind_{[\tau^j, \infty)}-\widetilde{\gamma}^j \centerdot \boldsymbol\lambda$. 
We consider separately the cases of continuous and purely discontinuous $P$. When $P$ is a continuous $(\mathbb{F},\mathbb{P})$ local martingale, $P$ is the $(\mathbb{F},\mathbb{P})$ stochastic integral, in the sense of local martingales, of an $\mathbb{F}$ predictable $(n+2)$ dimensional process $H=(H^k, k\in N)$ with respect to the $(n+2)$ dimensional process $(\hat{W}^k, k\in N)$. Since the matrix $(\frac{d\cro{\hat{W}^k,\hat{W}^{k'}}_{t}}{dt}, k,k'\in N)$ is uniformly positive-definite, for every $k\in N$, $H^k$ is individually $(\mathbb{F},\mathbb{P})$ integrable with respect to the one dimensional Brownian motion $\hat{W}^k$ in the sense of local martingales  (see \citeN[Chapter III, Section 4]{JacodShiryaev03}).
%Equivalently,  $(H^k)^2\centerdot\boldsymbol\lambda$ is finite $\mathbb{\P}$ a.s., hence $\mathbb{\Q}$ a.s.. Hence 
%$H^k$ is individually $(\mathbb{F},\mathbb{Q})$ integrable with respect to $\hat{W}^k$ in the sense of local martingales.
Moreover, as $\cro{H^k\centerdot\hat{W}^k,\nu}$ exists under
$\mathbb{P}$ (noting $H^k\centerdot\hat{W}^k$ is continuous),
 by the Girsanov theorem (see \citeN[Theorem 12.13]{HeWangYan92}),
 $\cro{H^k\centerdot\hat{W}^k,\nu}$ is of locally integrable total variation under $\mathbb{Q}$.
Hence
 $H^k$ is integrable with respect to  $\cro{\hat{W}^k,\nu}$ under $\mathbb{Q}$.
Moreover the $(\mathbb{F},\mathbb{Q})$ martingale part of $\hat{W}^k$ is an $(\mathbb{F},\mathbb{Q})$ Brownian motion, hence the $(\mathbb{F},\mathbb{Q})$ integrability of $H^k$ against this martingale part reduces to the a.s. finiteness of 
$(H^k)^2\centerdot\boldsymbol\lambda,$ which holds under $\mathbb{P}$ and therefore under $\mathbb{Q}$.
In sum, $H^k$ is $(\mathbb{F},\mathbb{Q})$ integrable with respect to  $\hat{W}^k$ in the sense of semimartingales. 
%Hence  $H^k$ is integrable with respect to  $\hat{W}^k$ under $\mathbb{Q}$ in the sense of semimartingales. 
As the hypothesis (H') holds between $\mathbb{F}\subset\mathbb{G}$ under $\mathbb{Q}$ (see after \qr{Ftsplit}), \citeN[Proposition 2.1]{Jeulin1980} implies that $H^k$ is $(\mathbb{G},\mathbb{Q})$ integrable with respect to  $\hat{W}^k$ in the sense of semimartingales. By \citeN[Lemma 2.1]{JeanblancSong12}, the stochastic integrals in the sense of the $(\mathbb{F},\mathbb{Q})$ semimartingales and in the sense of the $(\mathbb{G},\mathbb{Q})$ semimartingales are the same, hence
$P 
=
\sum_{k\in N} H^k\centerdot \hat{W}^k$ also holds in the sense of $(\mathbb{G},\mathbb{Q})$ semimartingales.
By Lemma \ref{beta-beta}, $(\hat{W}^k)^{\tau_{-1}\wedge\tau_{0}}=({W}^k)^{\tau_{-1}\wedge\tau_{0}}$ is a 
$(\mathbb{G},\mathbb{Q})$ local martingale.
 Applying \citeN[Theorem 9.16]{HeWangYan92}, 
%si int sto semima est est semima spéciale, alors 'la contradiction d'Emery ne peut se produire,', i.e. dès lors que intégrateur est ma loc, so is int sto semima, qui n'est autre que int sto ma loc
we conclude that, in fact, $H^k$ is $(\mathbb{G},\mathbb{Q})$ integrable with respect to  $(\hat{W}^k)^{\tau_{-1}\wedge\tau_{0}}$ in the sense of local martingales. 
Hence
$$P^{\tau_{-1}\wedge\tau_{0}-}
=
\sum_{k\in N}(H^k\centerdot \hat{W}^k)^{\tau_{-1}\wedge\tau_{0}-}
=
\sum_{k\in N}(H^k\centerdot {W}^k)^{\tau_{-1}\wedge\tau_{0}} 
$$
is a $(\mathbb{G},\mathbb{Q})$ local martingale.

Consider now the case of $P$ purely discontinuous. Without loss of generality we suppose that the locally bounded process $P$ is in fact bounded. Then, $P$ is the $(\mathbb{F},\mathbb{P})$ stochastic integral (in the sense of local martingale) of an $\mathbb{F}$ predictable $n$ dimensional process $K=(K^j, j\in N^\star)$ with respect to the $n$ dimensional process $(\hat{M}^j, j\in N^\star)$. The processes $\hat{M}^j, j\in N^\star,$ have disjoint jump times with jump amplitude 1. This implies that $K^j$ is integrable with respect to $\hat{M}^j$ individually. Moreover, as $P$ is bounded, the random variables $K_{\tau_{j}}, j\in N^\star,$ are bounded, hence $K$ itself is bounded (cf. \shortciteN[Theorem 7.23]{HeWangYan92}). As a consequence, $K$ is automatically $(\mathbb{G},\mathbb{Q})$ integrable with respect to $(\hat{M}^j, j\in N^\star)$ in the sense of local martingale. By \citeN[Lemma 2.1]{JeanblancSong12} again,
$$
P^{\tau_{-1}\wedge\tau_{0}-}
=
\sum_{j\in N^\star}(K^j\centerdot \hat{M}^j)^{\tau_{-1}\wedge\tau_{0}-}
=
\sum_{j\in N^\star}(K^j\centerdot {M}^j)^{\tau_{-1}\wedge\tau_{0}},
$$
which is a $(\mathbb{G},\mathbb{Q})$ local martingale.
%The invariance property of the probability measure $\mathbb{P}$ in regard to the quadruplet $(\tau_{-1}\wedge \tau_{0},\mathbb{F}, \mathbb{G}, \mathbb{Q})$ on the horizon $[0,T]$ for any constant $T>0$ is proved. 
\finproof

\subsection{Alternative Proof of the Condition (A)}\label{ss:suffcond}

Theorem \ref{DGC2} yields an explicit construction of the invariance probability measure $\mathbb{P}$ in the DGC model.
If we only want to establish the condition (A), i.e. the existence of $\mathbb{P}$,
a shorter proof is available based on the sufficiency condition of \citeN[Theorem 5.1]{CrepeySong15c}.

\bl
The $(\mathbb{G}, \mathbb{Q})$ intensity $\gamma$ of the random time $\tau_{-1}\wedge\tau_{0}$ is given by $$
\gamma = \ind_{[0,\tau_{-1}\wedge\tau_{0}]}(\gamma^{-1}+\gamma^0).
$$
\el

\proof
This follows from, for example, 
\citeN[Lemma 6.2]{CrepeySong15FS}.
\finproof

\begin{theorem}
The condition (A) holds in the DGC model $(\tau_{-1}\wedge \tau_{0},\mathbb{F}, \mathbb{G}, \mathbb{Q}).$ 
\end{theorem}

\proof Given a constant horizon $T>0$,
according to \citeN[Theorem 5.1]{CrepeySong15c},
we only need to prove the exponential integrability of
$\int_0^{\tau\wedge T} \gamma_s ds$, which can be done similarly to the proof of Lemma \ref{tt1}. 
\finproof

\section{Wrong Way Risk}

%This reflects a departure from the immersion property, whereby the default intensities of the surviving names and therefore the value of credit protection  
%spike at default times, in line with the wrong-way risk feature
%of counterparty risk embedded in credit derivatives, i.e.~the adverse dependence between the default risk of a counterparty and the underlying credit derivative exposure. 

%the wrong-way risk feature of the DGC model, namely, 
As visible in  \qr{QoverSci}, the default intensities
of the surviving names 
%and the value of the CDS protection 
spike at defaults
in the DGC model. This is very much related to the departure from the immersion property in this model, i.e. the fact that the invariance probability measure $\mathbb{P}$ is not equal to the pricing measure $\mathbb{Q}$ on $\cF_T.$ 
%As a consequence, the value of CDS protection spikes in the model, consistent with what is observed in the market. 
This 'wrong way risk' feature (cf. \citeN{CrepeySong15FS}) makes the DGC model appropriate for dealing with
counterparty risk on credit derivatives, notably portfolios of CDS contracts traded between a bank and its counterparty, respectively labeled as $-1$ and $0,$ and bearing on 
reference firms $i=1,\ldots,n.$

To illustrate this numerically, in this concluding section of the paper,
we study  the
valuation adjustment accounting for counterparty and funding risks (total valuation adjustment TVA)
embedded in one CDS 
%(credit default swap) 
between a bank and its counterparty on a third reference firm.

In Figure \ref{f:fig}, the left graph shows the TVA computed
as a function of the correlation parameter $\varrho$ in a DGC model of the three credit names (hence $n=1$): the bank, its counterparty and the reference credit name of the CDS. The different curves correspond to different levels of credit spread $\bar{\lambda}$ of bank: the higher $\bar{\lambda}$, the higher the funding costs for the bank, resulting in higher TVAs. 
All the TVA numbers are computed by
a
Monte Carlo scheme dubbed ``FT scheme of order 3'' in \citeN[Section 6.1]{CrepeyNguyen15}.
FT refers to
\citet*{FujiiTakahashi12a,FujiiTakahashi12b}.
%\citeANP{FujiiTakahashi12a} (\citeyearNP{FujiiTakahashi12a},\citeyearNP{FujiiTakahashi12b}). 
The numerical parameters  are set as in  \citeN[Section 6.1]{CrepeyNguyen15}, to which we refer the reader for a complete description of the CDS contract, of the FT numerical scheme and of other numerical experiments involving CDS portfolios (as opposed to a single contract here).

The right panel of Figure \ref{f:fig} shows the analog of the left graph, but in a fake DGC model, where we deliberately ignore the impact of the default of the counterparty in the valuation of the CDS at time $\tau_{-1}\wedge\tau_0$ (technically, in the notation of \citeN[Equation (6.7)]{CrepeySong15FS}, we
replace ($\tilde{P}^e_t+\tilde{\Delta}^e_t$) by $P_{t-}$ in the coefficient $\hat{f}$), in order to kill the wrong-way risk feature of the DGC model. We can see from the figure that, for large $\varrho,$ 
the corresponding fake TVA numbers are five to ten times  
smaller than the ``true'' TVA levels that can be seen in the left panel. In addition of being much smaller for large $\varrho,$  
the fake DGC TVA numbers in the right panel are mostly decreasing with $\varrho$. This shows
that the wrong-way risk feature of the DGC model is indeed responsible for the ``systemic'' increasing pattern observed in the left panel. 
\begin{figure}[htbp]
\begin{center}
\includegraphics[width=0.49\textwidth,height=0.39\textwidth]{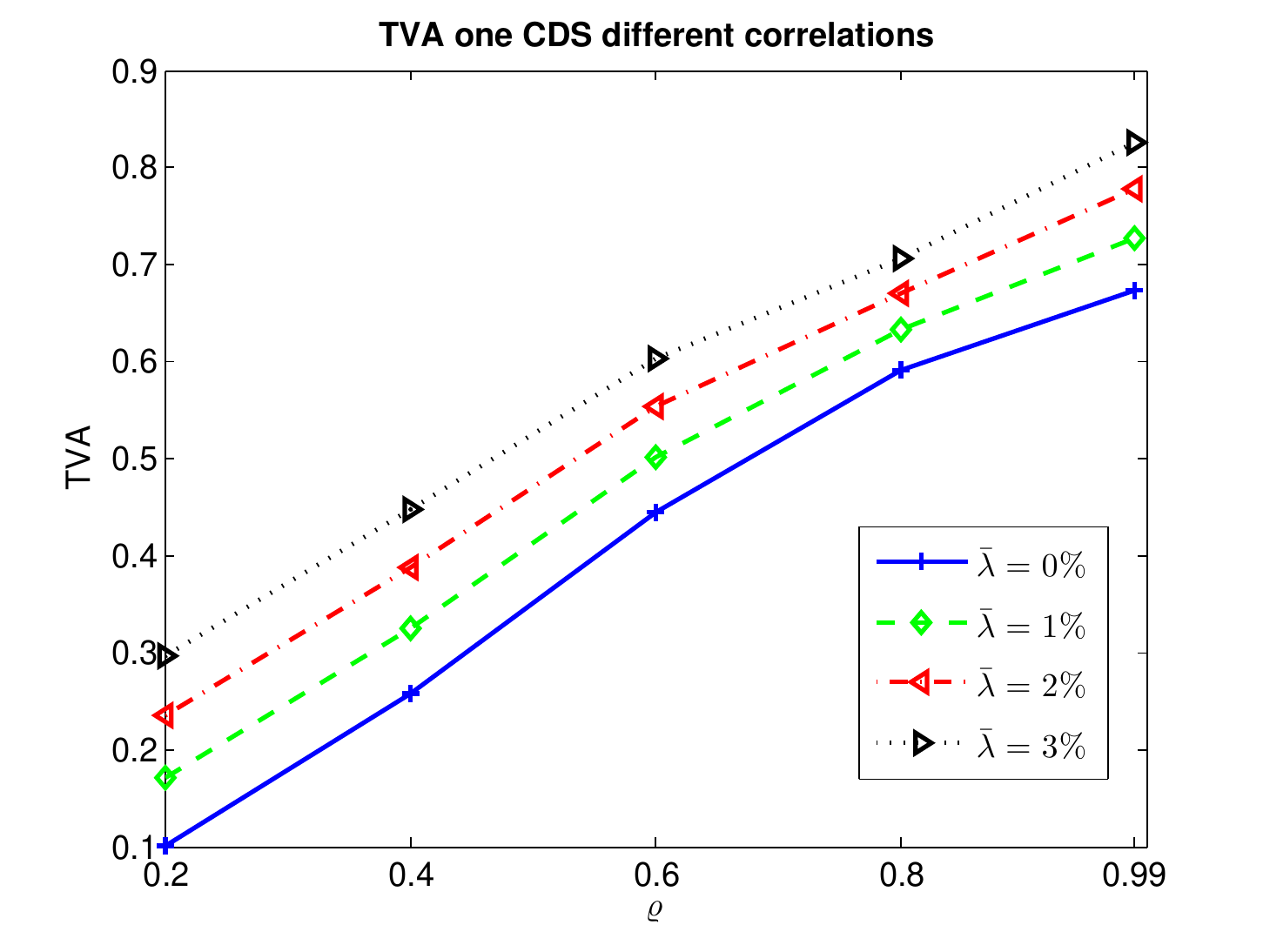} 
\includegraphics[width=0.49\textwidth,height=0.39\textwidth]{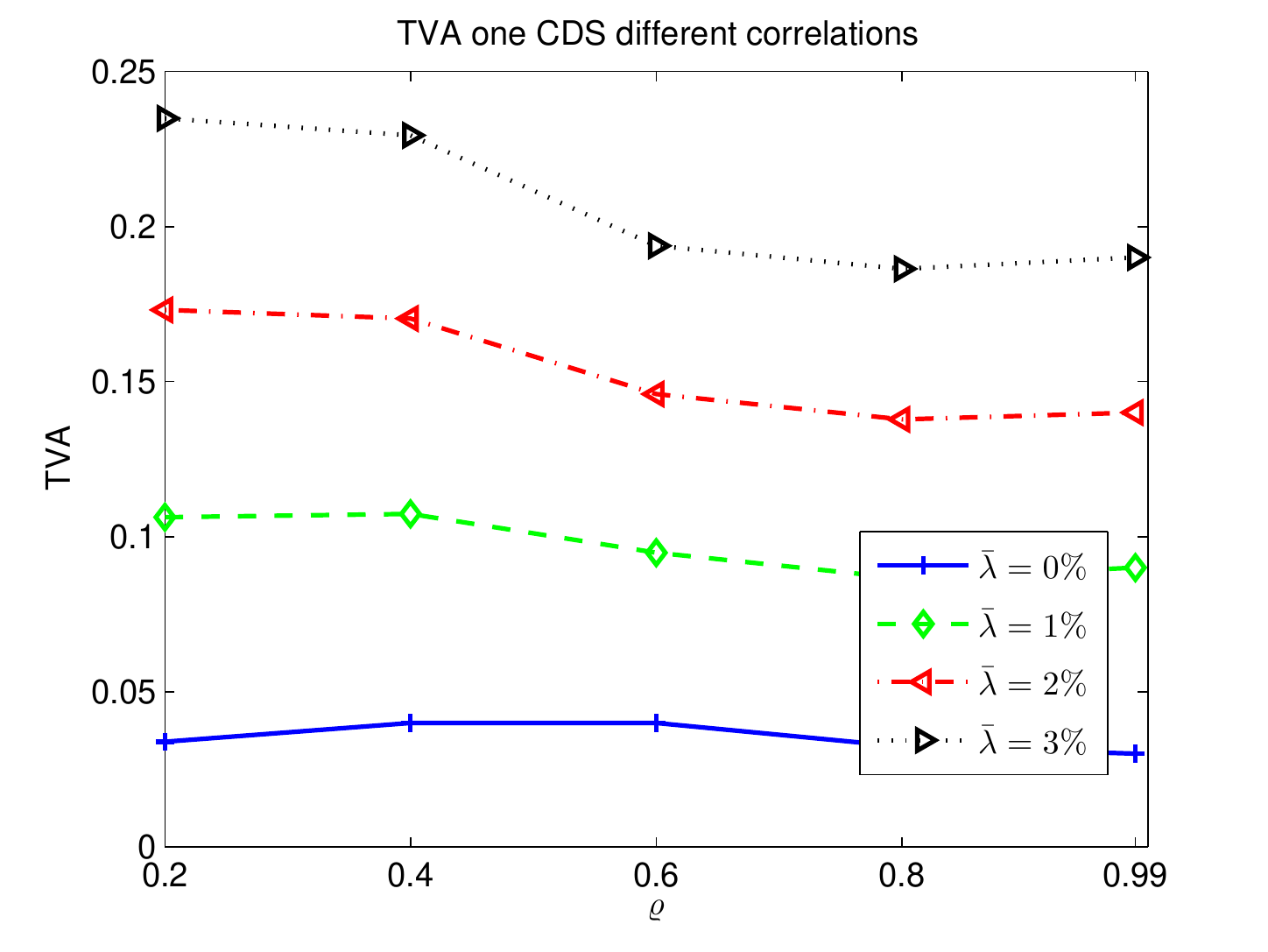}
\end{center}
\caption{{\it Left}: TVA on one CDS 
as a function of the correlation parameter $\varrho$ and for different bank credit spreads $\bar{\lambda},$
in a DGC model with three names: the bank, its counterparty and the reference credit name of the CDS.
{\it Right}: Analog results in a fake DGC model without wrong-way-risk.}
\label{f:fig}
\end{figure}
\appendix\normalsize

{
\def\bb{\mathbb{B}}
\def\B{\mathcal{B}}

\def\kI{\mbox{supp}(\bthet_t)}\def\kI{\mathfrak{I}}\def\kI{\mathcal{I}}
\def\kIs{\mbox{supp}(\tilde{\bthet}_t)}\def\kIs{\mathfrak{I}^\star}\def\kIs{\mathcal{I}^\star}
\def\BIs{\mathcal{B}^\star} 
\def\kJ{{\mathfrak{J}}}\def\kJ{{\mathcal{J}}} 
\def\kJs{\mathfrak{J}^\star}\def\kJs{\mathcal{J}^\star}
\def\kJt{\tilde{\mathfrak{J}}} \def\kJt{\tilde{\mathcal{J}}} 

\def\thexit{\hat{\xi}}\def\thexit{\xi^{\star}}

\def\bal{\begin{aligned}}
\def\eal{\end{aligned}}

\def\la{\label}
\def\bbf{\mathbf}
\def\al{\alpha}
\def\beq{\begin{eqnarray}}
\def\eeq{\end{eqnarray}}
\def\La{\theLambda}
\def\si{\sigma}\def\si{\mynu}
\def\sommen{\sum_{j,k\in V_N}}
\def\sommezd{\sum_{k\in\bbZ^d}}
\def\eps{\varepsilon}
\def\what{$\clubsuit\clubsuit\clubsuit$}
\def\grad{\bigtriangledown}
\def\demonstration{\noindent{\textbf{Proof. }}}
\def\sss{\subset\subset}
\def\trace{\,{\rm Tr}\,}
\def\Om{\Omega}
\def\om{\omega}
\def\gb{\overline{g}}
\def\gc{\hat{g}}

\def\PP{\mathbb P}\def\PP{\Proba}

\def\Ne{\N^{\ast}}
\def\CC{{\cal C}}
\def\DD{{\cal D}}
\def\B{\mathcal{B}}
\def\SSS{{\cal S}}
\def\E{{\mathbb E}}
\def\hh{{\mathbb H}}
\def\F{{\cal B}}\def\F{{\cal F}}
\def\G{{\cal G}}
\def\cH{{\cal H}}
\def\HHl{{\cal H}_{loc}}
\def\II{{\cal I}}
\def\JJ{{\cal J}}
\def\KK{{\cal K}}
\def\LL{{\cal L}}
\def\ft{\tilde{f}}
\def\tf{\tilde{f}}
\def\tth{\tilde{h}}
\def\tg{\tilde{g}}
\def\gt{\tilde{g}}
\def\et{\tilde{E}}
\def\ut{\tilde{u}}
\def\phit{\tilde{\Phi}}
\def\hf{\hat{f}}
\def\hg{\hat{g}}
\def\NN{{\cal N}}
\def\MM{{\cal M}}
\def\OO{{\cal O}}
\def\BP{{ \overline{{\cal P}}}}
\def\R{\mathbb{R}}
\def\D{\mathbb{D}}
\def\dis{\displaystyle}
\def\Z{\mathbb{Z}}
\def\C{\mathbb{C}}
\def\ind{\mathds{1}}
\def\rgt{\rightarrow}
\def\p{^{\prime}}
\def\vphi{\varphi}
\def\si{\varsigma}\def\si{\sigma}\def\si{\mynu}
\def\pa{\partial}
\def\para{\parallel}
\def\ds{D}
\def\gr{\sigma^{\ast}\nabla u}
\def\db{d{{B}}_s}
\def\dbt{d{B}_t}
\def\HD{F}
\def\disp{\displaystyle}
\def\Es{(E^{\ast})}
\def\dpr{d^{\prime}}
\def\hu{\hat{u}}
\def\ve{\varepsilon}
\def\un{ {\bf 1}}

\def\them{m}\def\them{\mathbf{m}}
\def\thel{l}
\def\myGt{G}
\def\thel{l}
\def\theCt{C}
\def\theS{\kappa}\def\theS{S}
\def\theSl{\kappa_l}\def\theSl{S_l}
\def\thekappa{\kappa}
%% dongli's macro
\def\hC{\hat{C}}
\def\tC{\tilde{C}}
\def\Q{\mathbb{Q}}
\def\tZ{\tilde{Z}}
\def\hZ{\hat{Z}}
\def\ff{\mathbb{B}}\def\ff{\mathbb{F}}
\def\ta{\tilde{a}}
\def\tb{\tilde{b}}
\def\ii{\imath}
\def\jj{\jmath}

\def\ir{\textcolor{red}}
\def\mr{\textcolor{green}}
\def\mb{\textcolor{blue}}
\def\bc#1{#1}\def\bc#1{\begin{color}{blue}#1\end{color}}
\def\rc{\textcolor{red}}
\def\r{\textcolor{red}}
\def\b{\textcolor{blue}}

\def\bal{\begin{aligned}}
\def\eal{\end{aligned}}
\def\thel{l}
\def\myGt{G}
\def\thel{l}
\def\theCt{C}
\def\theS{\kappa}\def\theS{S}
\def\theSl{\kappa_l}\def\theSl{S_l}

\def\hat{\widehat}

\def\thel{\ell}
\def\az{a} 
\def\thel{\ell}\def\thel{k}
\def\ind {1\!\!1}\def\ind{\mathds{1}}
\def\I{\ind}
\def \PP{{\mathbb Q}}
\def\theN{N}
\def\qqq{\quad\quad\quad}
\def\Phis{\Phi^\star}\def\Phis{\Phi}
\def\dl{\dot{\Psi}}\def\dl{\boldsymbol\psi^{i}}\def\dl{\psi^{i}}\def\dl{\psi^{j}}
\def\dli{\psi^{i}}
\def\z{\mathbf{z}}
\def\Z{\mathbf{Z}}
\def\cro#1{\langle #1\rangle}
\def\cZ{\mathcal{Z}}
\def\cZt{\tilde{\mathcal{Z}}}
\def\cZs{\mathcal{Z}^\star}
\def\Zm{\widetilde{m}}\def\Zm{\zeta}
\def\thel{l}
\def\thisc{c}
\def\theC{C}
\def\themu{\eta}\def\themu{\beta}
\def\mu{{\nu}}

\section{Gaussian Estimates}\label{a:prtt0}
\def\theg{g}

In this appendix we derive the Gaussian estimates that are used in the proofs of Lemmas \ref{corB} and \ref{tt1}.

\bl\label{EG}
Given a positive decreasing continuously differentiable function $\Gamma$ on $\mathbb{R}_+$ such that $$\int_{\mathbb{R}_+}t^d\Gamma(t)dt<\infty ~\mbox{ 
and }~\lim_{t\uparrow\infty}t^{d-1}\Gamma(t)\rightarrow 0,$$
for some integer $d\geq 0$, we write 
$
\theg (y)=-\frac{\Gamma'(y)}{\Gamma(y)}\sp G(y)=\int_y^\infty t^d\Gamma(t)dt .$
Let $\overline{y}\geq 0$ and $\alpha,\epsilon>0$. 
\hfill\break{\rm \textbf{{(i)}}} If $\theg (y)\geq \alpha y $ for $y>\overline{y}$,
then
$$
G(y)\leq \big(\frac{1}{\alpha}+\epsilon\big)y^{d-1}\Gamma(y) 
\mbox{ \quad for\quad  } y>\overline{y}\vee {\sqrt{|d-1| ({1\over{\epsilon\alpha^2}}+{1\over \alpha })}}.$$ 
\hfill\break{\rm \textbf{{(ii)}}} If $\theg (y)\leq 
\alpha
 y $ for $y>\overline{y}$,
then $$
G(y)\geq \big(\frac{1}{\alpha}
-\epsilon\big)y^{d-1}\Gamma(y) 
\mbox{  \quad for\quad } y>\overline{y}\vee {\sqrt{|d-1| ({1\over{\epsilon\alpha^2}}-{1\over \alpha })}}.$$
\el

\proof (i) For every positive continuously differentiable function $\varphi$ on $(0,+\infty)$,
\bel
&\left(G(y)-\varphi(y)\Gamma(y)\right)'
=
-y^d\Gamma(y)-\varphi'(y)\Gamma(y)+\varphi(y)\theg (y)\Gamma(y)\\
&\qqq=
(\varphi(y)\theg (y)-y^d-\varphi'(y))\Gamma(y) 
 \geq 
(\alpha y\varphi(y) -y^d-\varphi'(y))\Gamma(y) 
\eel
for $y\geq \overline{y}$. For $\varphi(y)=(\frac{1}{\alpha}+\epsilon)y^{d-1}$, 
\bel
&\alpha y\varphi(y) -y^d-\varphi'(y)=
(1+\epsilon\alpha)y^d -y^d-(\frac{1}{\alpha}+\epsilon)(d-1)y^{d-2} 
=\\ &\qqq
\epsilon\alpha y^d -(\frac{1}{\alpha}+\epsilon)(d-1)y^{d-2} 
=
(\epsilon\alpha y^2 -(\frac{1}{\alpha}+\epsilon)(d-1))y^{d-2}.
\eel
Therefore, if $y>\overline{y}\vee {\sqrt{|d-1| ({1\over{\epsilon\alpha^2}}+{1\over \alpha })}},$ 
then $\left(G(y)-\varphi(y)\Gamma(y)\right)'\geq \alpha y\varphi(y) -y^d-\varphi'(y)\geq 0.$
{But $\lim_{y\uparrow\infty}\left(G(y)- {\varphi(y)}\Gamma(y)\right)=0$, hence $G(y)-\varphi(y)\Gamma(y)\leq 0$. } 

\vspace{6pt}
\noindent
(ii) We again begin with
$$
\dcb
\left(G(y)-\varphi(y)\Gamma(y)\right)'
&=&
(\varphi(y)\theg (y)-y^k-\varphi'(y))\Gamma(y)\\
&\leq&
(\varphi(y)(\alpha y+\alpha')-y^k-\varphi'(y))\Gamma(y) 
\dce
$$
for $y\geq \overline{y}$.
We conclude as in (i) based on
$\varphi(y)=(\frac{1}{\alpha}-\epsilon)y^{k-1}$, assuming $\frac{1}{\alpha}>\epsilon$ (otherwise (ii) obviously holds).
% $$
%\dcb
%&&\varphi(y)(\beta y -y^k-\varphi'(y)\\
%&=&
%(\frac{1}{\beta}-\epsilon)y^{k-1}(\beta y+\beta')-y^k-(\frac{1}{\beta}-\epsilon)(k-1)y^{k-2}\\
%
%&=&
%(1-\epsilon\beta)y^k+(\frac{1}{\beta}-\epsilon)\beta'y^{k-1}-y^k-(\frac{1}{\beta}-\epsilon)(k-1)y^{k-2}\\
%
%&=&
%-\epsilon\beta y^k+(\frac{1}{\beta}-\epsilon)\beta'y^{k-1}-(\frac{1}{\beta}-\epsilon)(k-1)y^{k-2}\\
%
%&=&
%-(\epsilon\beta y^2-(\frac{1}{\beta}-\epsilon)\beta'y+(\frac{1}{\beta}-\epsilon)(k-1))y^{k-2}.
%\dce
%$$ 
%Note that, when $(\frac{1}{\beta}-\epsilon)^2\beta'^2-4\epsilon\beta(\frac{1}{\beta}-\epsilon)(k-1)>0$,$$
%\frac{(\frac{1}{\beta}-\epsilon)\beta'+\sqrt{(\frac{1}{\beta}-\epsilon)^2\beta'^2-4\epsilon\beta(\frac{1}{\beta}-\epsilon)(k-1)}}{2\epsilon\beta}
%\leq
%\frac{(\frac{1}{\beta}-\epsilon)\beta'}{\epsilon\beta}
%+
%\frac{\sqrt{\epsilon\beta(\frac{1}{\beta}-\epsilon)|k-1|}}{\epsilon\beta}.
%$$
%We conclude that, if $y$ is larger than the right hand side, $\varphi(y)(\beta y+\beta')-y^k-\varphi'(y)\leq 0$. {As $\lim_{y\uparrow\infty}\left(G(y)-\s{\varphi(y)}\Gamma(y)\right)=0$, this proves (ii).} 
\finproof\\

\noindent
We use the notation \qr{eqa initial enlargement} as well as
$\Phi$ and $\phi$
for the standard normal survival and density functions.
By a first application
of Lemma \ref{EG},
to the standard normal
density $\Gamma=\phi$, we recover the following classical inequalities on 
%the logarithmic density 
$\psi={\phi\over \Phi}$: 
for any constant {$\thisc >1$},
 %$\thisc '>1$,
\begin{equation}\label{e:basg}
{ \thisc ^{-1} y} \leq \psi (y)\leq {\thisc } y ,\ y>y_0,
\end{equation}
for some $y_0>0$ depending on $\thisc .$ The following estimate, where $\thisc $ and $y_0$ are as here, can be seen as a multivariate extension of
the right hand side inequality in \eqref{e:basg}.
%\ecor
%\proof
%We apply Lemma \ref{EG} to the functions $\phi(y)$ and $\Phi(y)$ on $y\in\mathbb{R}_+$. Let $\Gamma(y)=\phi$. We have $$
%\phi(y)=\int_y^\infty t\Gamma(t)dt,\ \ \Phi(y)=\int_y^\infty \Gamma(t)dt,
%$$ 
%and $\Gamma'(y)=-y\Gamma(y)$ which gives $\theg (y)=y$.
%\finproof
 
\bl\label{estim} There exist {constants
$a$ and $b$} such that, for every $j\in J,$ 
\beql{e:cqo}
& 0\leq\dl_{\rho,\sigma}\big(\z\big)\leq a+ b||\z||_{\infty}. 
%\\
%&
% \frac{1}{\sigma\sqrt{1-\rho}}\left(\frac{1}{\Phi(y_0)}+\frac{\thisc z_k}{\sigma\sqrt{1-\rho}}\right)
%+
%\frac{\thisc }{\sigma\sqrt{1-\rho}}\frac{\sigma\sqrt{\rho}}{\sigma\sqrt{1-\rho}}(\frac{1}{\sigma\sqrt{\rho}} ||\z||_{\infty}+ \frac{1}{\sigma\sqrt{\rho}}\sigma\sqrt{1-\rho}y_0+1+4
%\beta\s{y_1}) . 
\eeql
\el
\proof By conditional independence of the components of a multivariate Gaussian vector with homogeneous pairwise correlation $\varrho$,
we
have $ \Phi_{\rho,\sigma}\big(\z\big)=\int_{\mathbb{R}} \Gamma( y)dy$,
where
$\Gamma( y)=\prod_{\thel\in J } \Phi\left(\frac{z_l
+\sigma\sqrt{\rho}\,y}{\sigma\sqrt{1-\rho}}\right )\,\phi(y).$ 
Hence
%(see \citeN{Li}), 
\beql{eq:exc}
%& \Phi_{\rho,\sigma}\big(\z\big)=\int_{\mathbb{R}} \prod_{j\in J}
%\Phi {\left(\frac{z_j +\sigma\sqrt{\rho}\,y}{\sigma\sqrt{1-\rho}}\right )}\,\phi(y)dy\\ 
& \dl_{\rho,\sigma}\big(\z\big)=\frac{1}{\sigma\sqrt{1-\rho}}\int_{\mathbb{R}}
w_{\rho,\sigma}(\z,y)\psi\left(\frac{z_j
+\sigma\sqrt{\rho}\,y}{\sigma\sqrt{1-\rho}}\right ) dy,
\eeql
where
$w_{\rho,\sigma}(\z,y)= {{\Gamma( y)}\over{\Phi_{\rho,\sigma}\big(\z\big)}}.$
Straightforward computations yield
$$
\theg (t)=-\frac{\Gamma'(t)}{\Gamma(t)}= \sum_{ \thel \in J}\psi (\frac{z_{ \thel }+\sigma\sqrt{\rho}t}{\sigma\sqrt{1-\rho}})\frac{\sigma\sqrt{\rho}}{\sigma\sqrt{1-\rho}}+t \geq t,
$$
whereas
for $
t> \max_{ \thel \in J}\frac{1}{\sigma\sqrt{\rho}}(\sigma\sqrt{1-\rho}y_0-z_{ \thel })
$
and
$ 
t>\frac{1}{\sigma\sqrt{\rho}}\max_{ \thel \in J}z_{ \thel },
$
we have 
$$
\dcb
\theg (t) \leq 
 \sum_{ \thel \in J}c\frac{z_{ \thel }+\sigma\sqrt{\rho}t}{\sigma\sqrt{1-\rho}}\frac{\sigma\sqrt{\rho}}{\sigma\sqrt{1-\rho}}+t 

% \leq 
% \sum_{ \thel \in J}2c\frac{\sigma\sqrt{\rho}t}{\sigma\sqrt{1-\rho}}\frac{\sigma\sqrt{\rho}}{\sigma\sqrt{1-\rho}}+t 
\leq\bar{\alpha} t,

\dce
$$
with $\bar{\alpha} \eq \sum_{ \thel \in J}2c\frac{\sigma\sqrt{\rho}}{\sigma\sqrt{1-\rho}}\frac{\sigma\sqrt{\rho}}{\sigma\sqrt{1-\rho}}+1\geq 1.$
Applying Lemma \ref{EG}(i) with $d=1,\alpha=1$ and $\epsilon=1$, respectively
(ii) with $d=0,$
$
\alpha=\bar{\alpha} 
$
and
$\epsilon=\frac{1}{2\bar{\alpha}},$
yields
$$
\int_y^\infty t\Gamma(t)dt
\leq
2\Gamma(y)\sp y>0 \mbox{,\quad respectively } 
\int_y^\infty \Gamma(t)dt
\geq
\frac{1}{2\bar{\alpha} y}\Gamma(y)\sp y>\overline{y} \vee\frac{1}{\sqrt{\bar{\alpha}}},$$
where $ 
\overline{y} 
=
\frac{1}{\sigma\sqrt{\rho}}\max_{ \thel \in J}|z_{ \thel }|+ \frac{1}{\sigma\sqrt{\rho}}\sigma\sqrt{1-\rho}y_0.
$
Thus, setting
$
y_1= \overline{y} +1=\frac{1}{\sigma\sqrt{\rho}}\max_{ \thel \in J}|z_{ \thel }|+ \frac{1}{\sigma\sqrt{\rho}}\sigma\sqrt{1-\rho}y_0+1,
$
%for $y>\overline{y}^I \vee\frac{1}{\sqrt{\bar{\alpha}}}$,
%where $$
%\overline{y}^I
%=
%\frac{1}{\sigma\sqrt{\rho}}\max_{ \thel \in J}|z_{ \thel }|+ \frac{1}{\sigma\sqrt{\rho}}\sigma\sqrt{1-\rho}y_0.
%$$ 
%Set now$$
%y_1=\max_I\overline{y}^I +1=\frac{1}{\sigma\sqrt{\rho}}\max_{ \thel \in N}|z_{ \thel }|+ \frac{1}{\sigma\sqrt{\rho}}\sigma\sqrt{1-\rho}y_0+1.
%$$
\bel&\int_0^\infty t\Gamma(t)dt=\int_0^{y_1} t\Gamma(t)dt
+
\int_{y_1}^\infty t\Gamma(t)dt \leq y_1\int_0^{y_1} \Gamma(t)dt
+2\Gamma(y_1)\\&\qqq
\leq y_1\int_0^{y_1} \Gamma(t)dt
+4\bar{\alpha} {{y_1}} \int_{y_1}^\infty \Gamma(t)dt \leq (1+4\bar{\alpha})\int_{0}^\infty \Gamma(t)dt ,
\eel 
i.e.~
\beql{i:w} \int_0^\infty t w_{\rho,\sigma}(\z,t) dt \leq (1+4\bar{\alpha}) y_1.\eeql
Now, by \eqref{eq:exc} and the right hand side inequality in \eqref{e:basg}, 
\beql{partPhi} 
0\leq \sigma&\sqrt{1 -\rho} 
 \,\dl_{\rho,\sigma}\big(\z\big) 
\\ 
&\leq 
\int_{\mathbb{R}} \left( \frac{1}{\Phi(y_0)}\ind_{\{\frac{z_j+\sigma\sqrt{\rho}y}{\sigma\sqrt{1-\rho}}\leq y_0\}}+\thisc \frac{z_j+\sigma\sqrt{\rho}y}{\sigma\sqrt{1-\rho}}\ind_{\{\frac{z_j+\sigma\sqrt{\rho}y}{\sigma\sqrt{1-\rho}}>y_0\}}\right) w_{\rho,\sigma}(\z,y)dy\\
&=
 \left( \frac{1}{\Phi(y_0)}+\frac{\thisc z_j}{\sigma\sqrt{1-\rho}}\right) +
 \frac{\thisc \sigma\sqrt{\rho}}
{\sigma\sqrt{1-\rho}} \int_{\mathbb{R}}\ind_{\{\sigma\sqrt{\rho}y>\sigma\sqrt{1-\rho}y_0-z_j\}}y w_{\rho,\sigma}(\z,y)dy\\
&\leq 
 \left(\frac{1}{\Phi(y_0)}+\frac{\thisc z_j}{\sigma\sqrt{1-\rho}}\right)
 +
 \frac{\thisc \sigma\sqrt{\rho}}
{\sigma\sqrt{1-\rho}} {\int_0^\infty y w_{\rho,\sigma}(\z,y)dy,}
\eeql
%because
%$$
%\dcb
%&&\int_{\mathbb{R}}\ind_{\{\sigma\sqrt{\rho}y>\sigma\sqrt{1-\rho}y_0-z_k\}}yw_{\rho,\sigma}(\z,y)dy \leq 
%\int_0^\infty t w_{\rho,\sigma}(\z,t)dt .
%\dce
%$$
so that by substitution of \eqref{i:w} into (\ref{partPhi})
\bel
0\leq \sigma \sqrt{1 -\rho} 
 \,\dl_{\rho,\sigma}\big(\z\big) 
 \leq
 \left(\frac{1}{\Phi(y_0)}+\frac{\thisc z_j}{\sigma\sqrt{1-\rho}}\right)
+
 \frac{\thisc \sigma\sqrt{\rho}}
{\sigma\sqrt{1-\rho}} (1+4\bar{\alpha}) y_1 .~\ok 
\eel

\bl\label{bab} Let $m_t=\int_0^t \varsigma(s)dB_s,$ 
where $B$ is a a univariate standard Brownian motion and
$\varsigma$ is a square integrable %\continuous
 function with unit
$L^2$ norm.
For any constant $q>0,$ 
%and $i\in N,$ 
$
e^{q \sup_{0\leq s\leq t} m_s^2}
$
is 
%\t{\xcancel{$\mathbb{Q}$}}
integrable for sufficiently small $t$. 
\el

\proof {The process $(m_t)_{t\geq 0}$ is equal in law to a time changed 
Brownian motion $(W_{\bar{t}})_{t\geq 0},$ where $W$ is a a univariate standard Brownian motion and $\bar{t}=\int_0^{t} \varsigma ^2(s)ds$ goes to 0 with $t$. Thus, it suffices to show the result with $m$ replaced by $W$.}
Let $r_t$ be the density function of the law of $\sup_{0\leq s\leq t}|W_s|$ and let $R_t(y)=\int_y^\infty r_t(x)dx, y>0$, so that\beql{e:finito}
&\mathbb{E}[e^{q \sup_{0\leq s\leq t} W_s^2}]
=
\int_0^\infty e^{q y^2}r_t (y)dy 
=-[R_t (y)e^{q y^2}]_0^\infty+2q \int_0^\infty y R_t (y) e^{q y^2}dy 
\eeql
and (using the reflection principle of the Brownian motion)
$$
\dcb
R_t(y)
&=&\mathbb{Q}[\sup_{0\leq s\leq t}(W^+_s+W^-_s)>y] \leq \mathbb{Q}[\sup_{0\leq s\leq t}W^+_s>\frac{y}{2}]
+\mathbb{Q}[\sup_{0\leq s\leq t}W^-_s>\frac{y}{2}]\\&=&2\mathbb{Q}[\sup_{0\leq s\leq t}W_s>\frac{y}{2}]=2\mathbb{Q}[|W_t|>\frac{y}{2}]=2\mathbb{Q}[|W_1|>\frac{y}{2\sqrt{t}}]=4 \Phi(\frac{y}{2\sqrt{t}}),
%\int_{\frac{y}{2\sqrt{t}}}^\infty\phi(x)dx,
\dce
$$
where by the left hand side in \qr{e:basg}
$$%\int_{y/2}^\infty\phi_t(x)dx =\mathbb{Q}(W_1>\frac{y}{2\sqrt{t}})=
\Phi(\frac{y}{2\sqrt{t}})\frac{y}{2\sqrt{t}}\leq c \phi (\frac{y}{2\sqrt{t}})=\frac{c}{\sqrt{2\pi}}e^{-\frac{y^2}{8t}}\sp \frac{y}{2\sqrt{t}}>y_0.$$
%Hence, for some constant $C'$, $
%R_t(y )y\leq C'\phi (\frac{y}{2\sqrt{t}})$ for sufficiently large $y$. 
Therefore, for $\frac{1}{8t}>q$, both terms are finite in the right hand side of \eqref{e:finito}, 
which shows the result.~\ok

%\bibliographystyle{chicago}
%\bibliography{../../ref} 

%\end{itemize}
\end{document}